\documentclass[12pt,preprint]{aastex62}

\shorttitle{Systematics in the Cepheid Distance Scale}
\shortauthors{Madore \& Freedman}

\begin{document}


\title{\bf Systematics in the Cepheid and TRGB Distance Scales:\\Metallicity Sensitivity of the Wesenheit Leavitt Law}


\author{\bf Barry F. Madore}
\affil{The Observatories \\ Carnegie Institution for Science \\ 813
Santa Barbara St., Pasadena, CA ~~91101}
\affil{Dept. of Astronomy \& Astrophysics, \\University of Chicago, Chicago, IL}
\email{barry.f.madore@gmail.com}
\author{\bf Wendy L. Freedman}
\affil{Dept. of Astronomy \& Astrophysics \& The Kavli Institute for Cosmological Physics, \\University of Chicago, Chicago, IL}
\email{wfreedman@uchicago.edu}

\begin{abstract}

Using an updated and significantly augmented sample of Cepheid and TRGB distances to 
28 nearby spiral and irregular galaxies, covering a wide range of metallicities, we have searched
for evidence of a correlation of the zero-point of the Cepheid Period-Luminosity relation 
with HII region (gas-phase) metallicities.   Our analysis, for the 21 galaxies closer than 12.5~Mpc, results in the following conclusions: 
 (1) The zero points of the Cepheid and TRGB distance scales are in remarkably good agreement, with the mean offset in the zero points of the most nearby distance-selected sample being close to zero, $\Delta\mu_o$(Cepheid - TRGB) = -- 0.026 $\pm$~0.015~mag (for an $I$-band TRGB zero point of $M_I = $ -- 4.05~mag); however, for the more distant sample, there is a larger offset between the two distance scales, amounting to -- 0.073 $\pm$~0.057~mag. (2) The individual differences, about that mean, have a measured scatter of $\pm$0.068~mag. (3) We find no statistically significant evidence for a metallicity dependence in the Cepheid distance scale using the reddening-free W(V,VI) period-luminosity relation: 

$\Delta \mu_o (Cepheid - TRGB) = - 0.022 ~(\pm0.015) \times ([O/H]-8.50) - 0.003 ~(\pm0.007)$

\end{abstract}


\section{Introduction}

Cepheids have been used for more than a century in probing the scale size of the Universe, beginning most notably with Henrietta Leavitt's understated discovery of  {\it``... a simple relation between the brightness of the variables ...}" (in the Small Magellanic Cloud) {\it ``... and their periods}" (Leavitt \& Pickering 1912). What we now understand about Cepheids and their Period-Luminosity-Color relation has been guided by theory, and  tested by methodical observational programs, involving calibration  at the limit of the available detectors and telescopes. Much progress has been made by taking advantage of strides in detector technology  (see for instance, McGonegal et al. 1982 and Freedman et al. 2012 for the impact of the introduction of near- and mid-infrared detectors.)
The scatter in the PL relation is now better understood, where 
known contributing factors include the intrinsic color (Sandage, 1972) and extrinsic reddening (both total line-of sight extinction and differential reddening). Metallicity effects can also, in principle, change the colors of individual Cepheids through wavelength-dependent differential line blanketing and re-radiation in their atmospheres (Breuval et al. 2021, 2022 and Owens et al. 2022).  Changes to a Cepheids' interior stellar structure, radii, luminosities and pulsation properties, are also manifest in systematic changes to the finite width, position and slope of the Cepheid instability strip as a whole (see Madore \& Freedman 1991, updated in Freedman \& Madore 2023). 

Thirty years ago, in their paper illustrating the promise of the Tip of the Red Giant Branch (TRGB) method 
as a distance indicator for resolved galaxies, Lee et al. (1993) reported two things 
of note in the context of this current study:
(1) The two independent distance scales (Cepheids and TRGB) {\it agreed systematically 
in zero point} at the 4\% level of accuracy. 
(2) The individual differences in the distances to the six galaxies in their study, that also had published HII-region metallicities, {\it showed no trend with metallicity} (upper-right panel of their Figure 5). 
The paper was exploratory, the samples were small, and all calibrations and tests for higher-order correlations were still in their infancy.  
Nevertheless, that study provided a novel, differential, distance-independent test for a metallicity dependence of the Cepheid PL relation.  

The TRGB method uses Population II red-giant branch (RGB) stars found in the dust-free, outer halos of most galaxies, regardless of Hubble type. 
In the $I$ band the TRGB is seen as a sharp discontinuity in the marginalized luminosity function of the RGB population (da Costa \& Armandroff 1990; Lee, Freedman \& Madore 1993). 
Metallicity variations within these intrinsically low-metallicity populations manifest themselves through atmospheric line-blanketing effects that shift the color of the tip of the RGB to the red with increasing metallicity, but with only a slight, color-correlated  decrease in the luminosity of the tip, that is well calibrated (e.g., Rizzi et al. 2007, Freedman 2021, Hoyt 2022) as measured in the $I$ band.
As such the $I$-band luminosity of the TRGB discontinuity has proven itself to be an excellent, easily identified, high-precision, and now widely used 
distance indicator.

Importantly for this comparison, these two distance indicators (Cepheids and TRGB stars) are not only spatially independent of each other (one confined to the disk, the other, which can be specifically targeted in the halo), but they also have very few systematics in common in their calibration and/or in their application, when it comes to determining distances \footnote{The dominant shared systematic is the geometric distances to the LMC and the maser galaxy NGC 4258 which are used to set the absolute magnitude zero point for both the Cepheids and the TRGB. And, the same foreground (Milky Way) reddenings are used for both the TRGB and Cepheid extinctions. But it has to be emphasized that since we are dealing with the differences between these two methods for each galaxy those shared systematics largely cancel.}. 
With a predetermined choice of fields, the TRGB method is optimally applied to old, low-mass, low-metallicity stars in the low-density (widely separated and uncrowded), dust- and gas-free halos of galaxies. 
The Cepheid PL relation is applied to young, high-mass, medium- to high-metallicity stars in the densely-populated (crowded), dusty, gas-rich disks of galaxies. For a more detailed discussion and history of the TRGB method see the introduction to the paper by Madore et al. (2023). 

Having two high-precision distance estimates to the same galaxies allows one to undertake direct, differential tests for systematic differences between the two methods. 
The first, and most obvious test is to look for any first-order offsets between the distances being measured by one method as compared (on a galaxy-by-galaxy basis) to distances being measured by the second method. 
Other tests can seek out higher-order correlations of these same differences as a function of other physical variables that might conceivably influence the luminosities of the stars involved. 
Some obvious parameters worth considering are: 
(a) the host galaxy type, 
(b) the  absolute magnitude (mass) of the host galaxy, 
(c) the metallicity of disk gas, out of which the Cepheids were recently formed, 
(d) the surface brightness of the disk in which the Cepheids are found, and 
(e) the distance to the host galaxy, where its surface brightness can act as a proxy for crowding and confusion. 
Correlations of this sort were sought out in the first paper on this topic by Lee, Freedman \& Madore (1993); and, as alluded to above, no significant correlations with any of these properties were found at that time.

It is not our intention here to review all of the varied observational and theoretical tests that have been proposed and/or undertaken in attempting to find and calibrate a metallicity sensitivity of the Cepheid PL relation, but it needs to be noted that, while there have been many attempts and many calibrations, no definite consensus has been reached on the magnitude, or even the sign, of the effect. For instance, contrary to most of the negative empirical values being advocated, Romaniello et al. (2006) and Bono et al. (2008) published positive correlations. Coefficients of the simply parameterized trend $\Delta$(mag) [Cepheid] = $\gamma~ \times$ [O/H] have ranged from declarations of a null detection (i.e., $\gamma$ = 0.0 mag/dex) by Udalski et al. 2001 and again by Freedman \& Madore 2011, and then similarly asserted more recently by Wielgorski et al. (2017); to moderate sensitivities of $\gamma \sim -0.2$ mag/dex (Freedman \& Madore 1990; Kennicutt et al. 1998; Sakai et al. 2004; Gieren et al. 2018; Breuval et al. 2021; Owens et al. 2022); and on up to significantly larger values, in the range of $\gamma = $ --0.4 to --0.5~mag/dex (Efstathiou 2014; Clementini et al., 2021).

Settling on a value for the metallicity sensitivity of the Cepheid PL relation has wider implications beyond the physics of the Cepheid PL relation itself. 
As calibrated by Cepheids, the derived value of the Hubble constant is dependent upon the degree of metallicity sensitivity of the Cepheid PL relation. 
As shown by Freedman et al. (2001) more than 20 years ago (and  demonstrated most recently by Efstathiou 2014 and 2020), in going from no metallicity corrections at all, to applying a steep dependence, such as his value of $\gamma = $ --0.53~mag/dex say, the derived value of the Hubble constant can drop from 73~km/s/Mpc to 67~km/s/Mpc. 
Compared to the ``factor-of-two controversy'', that raged at the end of the last century, this 10\% difference might not seem so consequential, were it not for the recent (and totally independent) results on the Hubble constant inferred from modeling the cosmic microwave background radiation, obtained by the Planck satellite (Ade et al. 2016, Aghanim et al. 2020).
The resulting ``tension'' between the very high-precision Planck value of $H_o = $ 67.4 $\pm$0.5 ~km/s/Mpc (Aghanim et al.) and the more recent locally-determined values of $H_o \sim $ ~73~km/s/Mpc (using Cepheids: Freedman et al. 2012, Riess et al. 2022) and $H_o \sim $ ~70~km/s/Mpc (using the TRGB Method: Freedman et al. 2019, 2020, 2021) now have implications not only for cosmology, but also for fundamental physics (see Freedman 2017, Verde et al. 2019 and references therein, for recent commentaries).
In this paper, we revisit  the question of the metallicity sensitivity of the Leavitt Law, based upon updated Cepheid and TRGB distances. Our goal is simply to update the considerable amount of new data since the study of Lee et al. (1993) 30 years ago, both for Cepheids and the TRGB.

\begin{figure*} 
\centering 
\includegraphics[width=18.0cm, angle=-0]{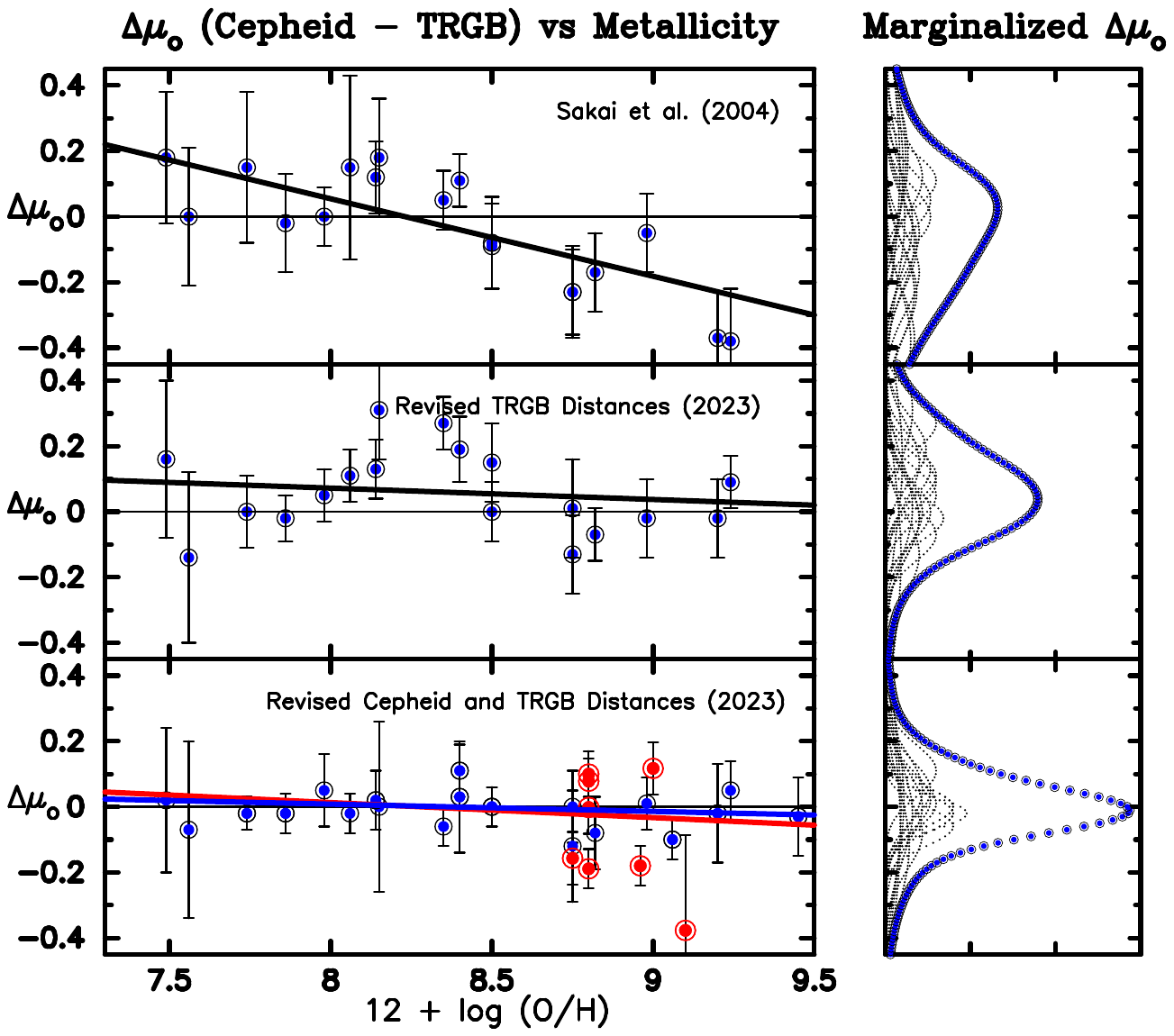} 
\caption{\small The decrease over time of the apparent correlation  of the Cepheid PL relation zero point with increasing metallicity. 
The top panel shows the data (in blue) replotted from Sakai et al. (2004), where an apparently strong negative correlation was reported. The middle plot recapitulates the Sakai plot, but only updating the TRGB moduli to their modern values. The bottom plot shows the results of updating both the Cepheid and TRGB distances. The blue line represents the fit to the blue (nearby sample) data points; the red line is the fit for the total sample, including the most distant galaxies (red data points). The right row shows the marginalized cumulative probability density distribution (in blue) for the nearest-galaxy sample (in blue) in the left panels.
Shallower (black) dotted are the individually differenced moduli shown as unit-area Gaussians. Galaxies with distance moduli exceeding 30.5~mag (as partitioned off in Figure 3) are highlighted as larger red dots, showing once again that the most distant galaxies are preferentially also among the highest-metallicity galaxies in this sample.}
\end{figure*}

\begin{figure*} 
\includegraphics[height=17.0cm,angle=-0]{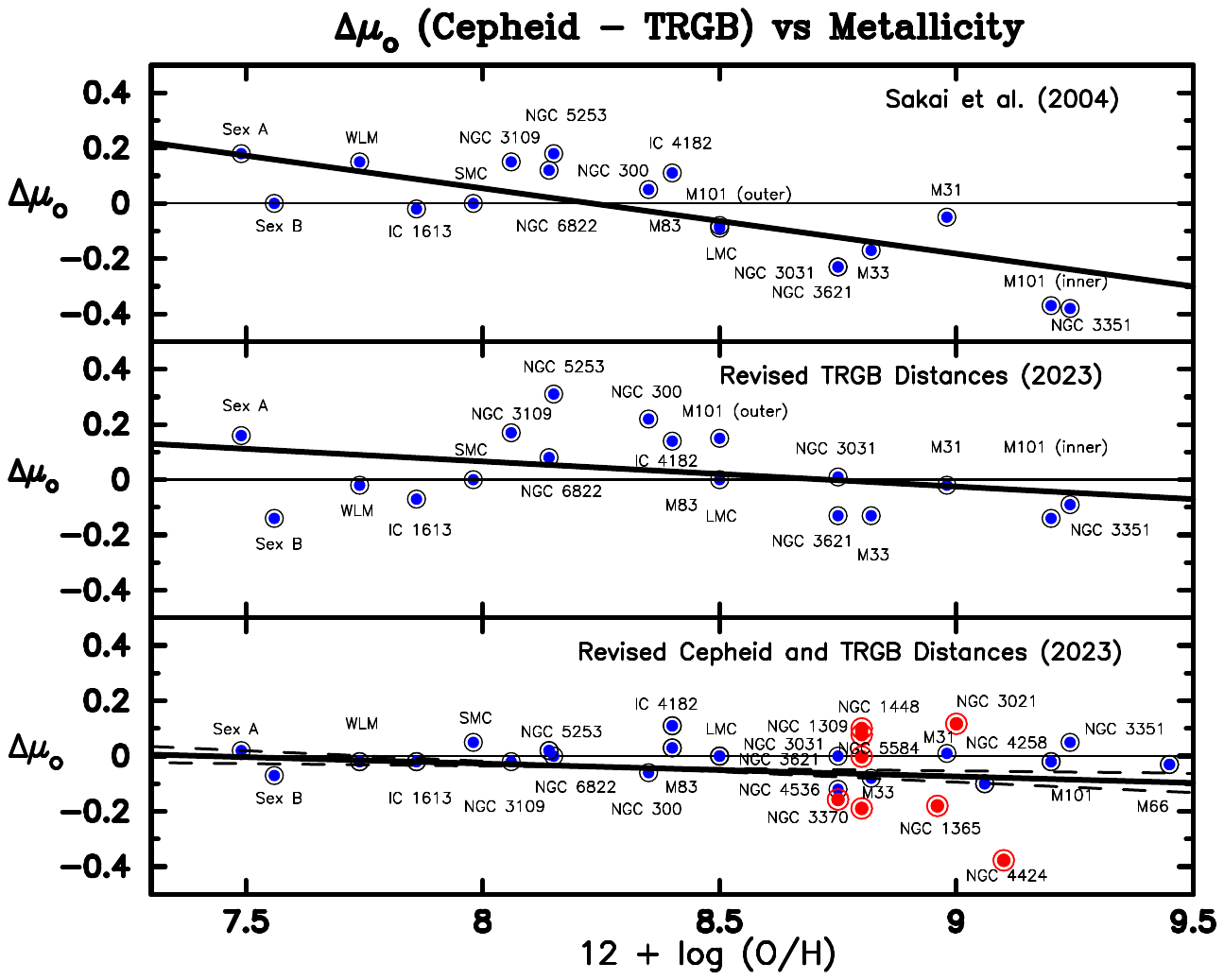} 
\caption{\small Same as Figure 1 but with galaxies individually identified.}
\end{figure*}
\eject
\begin{figure*} 
\includegraphics[width=18.0cm, angle=-0]
{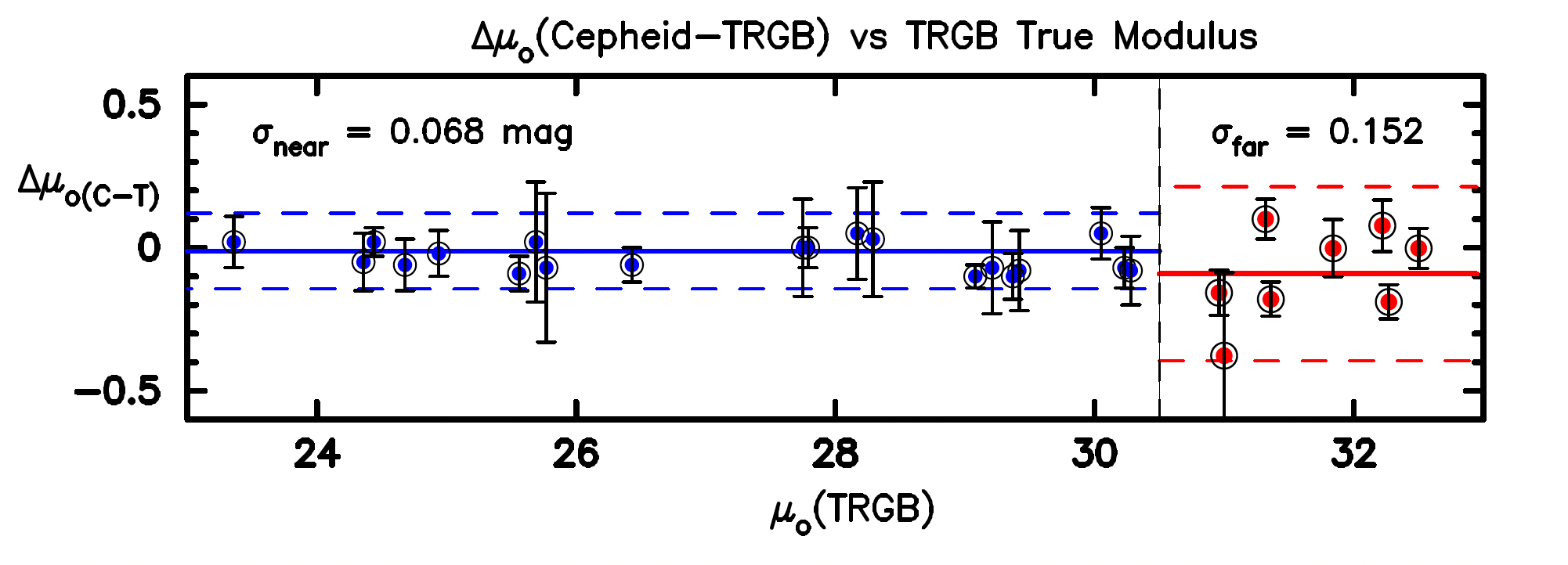}
\caption{\small Differences between the Cepheid and TRGB distance moduli as 
a function of distance.
The abrupt increase in scatter is marked by the broken vertical line at 
$\mu_o = $ 30.5~mag. 
At nearer distances the scatter is extremely small (at the level of
$\pm$0.068~mag) while 
beyond 12.5~Mpc the scatter is found to be at a level that is nearly a factor of two and a half times larger ($\pm$0.152~mag). The average error in the differences in distance moduli for the two (near and far) samples are 0.100 and 0.103 mag, respectively, which when compared to the scatter of the data points around the mean (0.068 and 0.152 mag, respectively), suggests that the errors on the nearby galaxies moduli are on average overestimated by about 47\% (0.032 mag), and the errors on the distant sample are underestimated by about 48\% (0.049 mag).
Solid horizontal lines show the mean offset from zero. Dashed horizontal lines are two-sigma bounds on the scatter in each of the two samples. Note that the more-distant sample of galaxies has both larger scatter in the sample, and systematically lower Cepheid distance moduli than their corresponding TRGB distance moduli.}
\end{figure*}
\section{The TRGB vs Cepheid Distance Test Revisited}
A search of the recent literature reveals that 28 nearby galaxies have had 
both TRGB and Cepheid distances measured to them. 
This is nearly twice as many galaxies as used, for example by Kennicutt et al. (1998) or Sakai et al. (2004), where this differential test was most recently undertaken 20 years ago. 
Moreover, many of those originally considered galaxies now have higher-precision data taken for them, both in terms of revised Cepheid distances, and also in much improved and homogeneously reduced TRGB distances (e.g., Tully et al. 2013, 2015) with a consistently adopted zero point of $M_I = $ ~-4.05~mag (Freedman et al. 2019, 2020). The Cepheid distances used here are derived uniformly from a W  = V - R(V-I) (Madore 1982), reddening-free version of the period-luminosity relation, the Wesenheit function. This function is expanded upon more in Section 3, and a full description of this form of the PL relation and its history is given in the Appendix.
These galaxies are listed in Table 1,  where we give the host galaxy name, the TRGB distance modulus and its reference, followed by the Cepheid distance modulus and its reference. 
The last two columns contain the mean metallicity for the Cepheids in that galaxy and its primary reference. 
When multiple values were reported for any given method, preferential consideration was given to determinations that had the highest quoted internal precision, those that were most recently published, and/or those that were derived from homogeneous compilations. 
In many cases the adopted value is also (coincidentally) close to the median of the totality of values published to date. 

\vfill\eject
\begin{deluxetable*}{lccccccccrc}
\tablecaption{TRGB - Cepheid Metallicity-Sensitivity Calibration Sample \label{tab:trgb_calibrators}} 
\tablehead{\colhead{Galaxy} & \colhead{$\mu$(TRGB)} &\colhead{err.}& \colhead{Ref} & \colhead{$\mu$(Cepheid)}& \colhead{err.} & \colhead{Ref} & \colhead{Z} & \colhead{Ref} }
\startdata
LMC    & 18.48  &0.040& 0 & 18.48 & 0.040& 0 & 8.50  & 4  \\
SMC    & 18.99  &0.08& 29 & 19.04 &0.08& 47 & 7.98  & 4    \\
NGC 0224 = M31 & 24.46  &0.07& 12 & 24.47 &0.05& 11 & 8.98  & 4   \\
NGC 0300 & 26.43  &0.07& 3 & 26.37 &0.09& 14,15 & 8.35  & 4   \\
NGC 0598  = M33 & 24.71&0.04& 7 & 24.62 &0.09& 9 & 8.82  & 4  \\
NGC 1309 & 32.50  &0.07& 18 & 32.50 &0.04& 39 & 8.80  & 3  \\
NGC 1365 & 31.36  &0.05& 18 & 31.18&0.05& 39 & 8.96  & 3  \\
NGC 1448 & 31.32  &0.06& 18 & 31.42 &0.02& 39 & 8.80  & 3  \\
NGC 3021 & 32.18  &0.03& 35 & 32.30 &0.07& 39 & 9.00  & 3  \\
NGC 3031 = M81 & 27.79  &0.07& 3,17 & 27.79 &0.09& 3,16 & 8.75 & 4\\
NGC 3109 & 25.56  &0.05& 7 & 25.54&0.03& 46 & 8.06  & 4   \\
NGC 3351 & 29.92  &0.05& 3 & 30.01 &0.08& 21 & 9.24  & 4   \\
NGC 3370 & 32.27  &0.05& 18 & 32.08 &0.04& 39 & 8.80  & 3  \\
NGC 3621 & 29.26  &0.12& 7 & 29.14 &0.06& 3 & 8.75  & 4   \\
NGC 3627 = M66 & 30.18  &0.09& 1 & 30.15&0.08& 2 & 9.25  & 21   \\
*NGC 4258 & 29.37  &0.02& 1 & 29.27 &0.06& 39 & 9.06  & 3  \\
NGC 4424 & 31.00  &0.06& 18 & 30.62 &0.28& 39 & 9.10  & 3  \\
NGC 4536 & 30.96 &0.05& 18 & 30.80 &0.06 & 39 & 8.75  & 3  \\
NGC 5236 = M83 & 28.29 &0.11& 3 & 28.32 &0.13& 5 & 8.40   & 6    \\
NGC 5253 & 27.75  &0.11& 3,19 & 27.75 &0.24& 3 & 8.15  & 4   \\
NGC 5457 = M101 & 29.07  &0.04& 41 & 29.05 &0.14& 20 & 9.20  & 40  \\
NGC 5584 & 31.76  &0.05& 1 & 31.84 &0.03& 39 & 8.80  & 3  \\
NGC 6822 & 23.36  &0.07& 3 & 23.38 &0.05& 25 & 8.50  & 4  \\
IC 1613 & 24.31  &0.05& 28 & 24.29 &0.03& 26 & 7.86  & 4   \\
IC 4182 & 28.17  &0.05& 3,7 & 28.22 &0.06& 45 & 8.40  & 4   \\
Sextans A  & 25.69  &0.06 & 44 & 25.71 &0.20& 27 & 7.49  & 4     \\
Sextans B  & 25.77 &0.03 & 10 & 25.69 &0.27& 28 & 7.56  & 4     \\
WLM & 24.94  & 0.03 & 42 & 24.92 & 0.04 & 41 & 7.74  & 4   \\
\enddata
{\tiny
References to Table 1:  
0: Fiducial: Pietrzynski et al. (2019);
1: Jang \& Lee (2017);  
2: Gibson et al. (2000)
3: Tully et al. (2013)
4: Sakai et al. (2004); 
5: Saha et al. (2006); 
6: Bresolin et al. (2009)
7: Rizzi et al. (2007); 
8: Dolphin \& Kennicutt (2002) 
9: Madore \& Freedman (1991);
10: Jacobs et al. (2009);
11: Kochanek et al. (1997);
12: Conn et al. (2012);
13: Gieren et al. (2013);
14: Bono et al. (2010);
15: Gieren et al.(2005);
16: Kanbur et al. (2003);
17: Radburn-Smith et al. (2011);
18: Freedman et al. (2019);
19: Tully et al. (2015);
20: Stetson et al. (1998);
21: Ferraresse et al. (2000);
22: Kennicutt et al. (1998);
24: Beaton et al. (2019);
23: Sabbi et al. (2018);
25: Rich et al. (2014); 
26: Scowcroft et al. (2013);
27: Pioto et al. (1994); 
28: Sakai et al. (1997); 
29: Cioni et al. (2000);
30: Hoyt et al. (2019);
31: Gibson et al. (1999);
32: Hatt et al. (2018a);
33: Zaritsky et al. (1999);
34: Hatt et al. (2018b);
35: Jang et al. (2018);
36: Jang \& Lee (2018);
37: Madore \& Freedman (2020);
38: Riess et al. (2011);
39: Table 2, this paper;
40: Mager et al. (2013);
41: Beaton et al. (2019);
42: McQuinn et al. (2017);
43: Gieren et al. (2008);
44: Dolphin et al. (2003);
45: Freedman et al. (2001);
46: Pietryzynski et al. (2006);
47: Marconi et al. (2017).}
\end{deluxetable*}
\vfill\eject
\medskip
Figures 1 and 2 contain our results in graphical form:
\medskip\par\noindent
(1) The upper panel reproduces the plot of differential distance moduli (in the sense Cepheid minus TRGB) as a function of host-galaxy HII-region abundances, as originally provided in Sakai et al. (2004). 
The only difference here is that, in these plots, we identify the individual galaxies so that one can more easily track changes that occur in the plots following below it. 
The solid black line is the originally published fit to the data, where a metallicity dependence with a slope of $\gamma = -0.24$ $\pm$ 0.05 mag/dex was reported. The scatter around the fit is found to be $\pm$0.12~mag.
 
\medskip\par\noindent
(2) The middle panel shows the incremental effect, on the scatter and on the trend with metallicity, caused by updating to the most recently published 
TRGB distances for the same sample of galaxies discussed by Sakai et al. (2014). 
The result of this first step is already quite dramatic:  In this updated version the slope of the relation has decreased by about a factor of six (from --0.24 to --0.04 mag/dex); while the scatter is unchanged, at $\pm$0.12 mag
around the regression.

\medskip\par\noindent
(3) The lower panel includes two additional changes: 
\par\noindent(a) An up-dating of the TRGB and Cepheid distances, and 
(b) The augmentation of the sample of galaxies entering the test, going from 17 to 29 systems, which are also listed in Table 1. 
 {\bf In this final panel of Figure 1,  there is now no evidence for any 
significant dependence of the Cepheid PL relation on metallicity.} The relation with metallicity is $\Delta \mu_o (Cepheid - TRGB) = -0.046 ~(\pm0.019) \times ([O/H]-8.50) - 0.010 ~(\pm0.011) $ for the entire sample of 29 galaxies, 
and is statistically flat $\Delta \mu_o (Cepheid - TRGB) = - 0.022 ~(\pm0.015) \times ([O/H]-8.50) - 0.003 ~(\pm0.007) $ for the subset of the 21 nearest galaxies having distance moduli less than 30.5~mag. Two of the lowest metallicity galaxies (Sextans A and B) are also two of the systems that have the largest uncertainties on their distances. The plotted solutions are weighted fits so the impact of these two galaxies on the final solution appropriately down-weighted and taken into account. The scatter in this final plot is $\pm$ 0.068~mag for the nearest sample, and is discussed in more detail in Figure 2.

\par\noindent

There are several factors driving the change from the earlier plots. First, as can be seen in Figure 1, most of the suggestion of a gradient in Sakai et al. (2004) resulted from only two data points falling at the highest metallicities, M101 and NGC~3351. The trend was considerably weaker for galaxies with metallicities 12 + log(O/H) $<$ 8.5 dex. In the intervening time, improvements to both the TRGB and Cepheid distances have served to reduce the uncertainties in each method, and data for many more galaxies with larger abundances have been obtained.

\section{Broader Implications and Extensions}

If, as concluded above, the specific combination of V and I bands in the form of W, is insensitive to metallicity, it then follows that V and I are well-suited for estimating unbiased values of extinction and color excess (assuming, of course, that the interstellar extinction law, extrapolated beyond the VI wavelength range to the red, is universal.)

Macri et al. (2001) undertook the first direct test of this conjecture when the near-infrared camera, NICMOS was installed on HST. They obtained H-band imaging of 70 Cepheids in 12 nearby galaxies chosen from the HST Key Project, asking the simple question: {\it For each of the galaxies, does an extrapolation of the standard Milky Way extinction curve, previously fit only to the VI Cepheid data, also fit the H-band data for those same galaxies and their Cepheids? }
The answer was, that, to within the quoted uncertainties, the adopted interstellar extinction curve is indeed universal.\footnote{ A few years later, this same conclusion was independently reached by Sakai et al. (2004) where they state ``{\it For some of the galaxies with ground-based distances, Cepheids are observed in more bands than V and I (for instance B and R, see Appendix A). Calculating a distance using multi-wavelength data sometimes leads to improvements over fits which only use V and I data, especially in the case of sparsely sampled PL relations. Distances using all available photometric bands are there fore listed in Table 3, Columns 5 and 6;} {\bf these agree identically to the distances tabulated in Columns 3 and 4, when only V and I data are available.} (emphasis ours). } Interestingly, this conclusion implicitly validates using VI data (without the need for additional NIR imaging) to obtain reddening-corrected true distance moduli to extragalactic samples of Cepheids.

We now ask: Do the H-band observations of the Cepheids studied in the SHoES project (Riess et al. 2016) carry any sensitivity to the metallicity of those stars? SHoES was put together as a Cepheid-based Type~Ia supernova calibration project, with the specific goal of addressing potential systematics in the Hubble constant, and reducing the combined (systematic and statistical) uncertainties in it final error budget. They used Cepheids, pushing the calibration into the near-infrared, specifically introducing H-band observations and combining them with VI data.
In introducing the SHoES sample into our analysis, we have to be careful to take the new Cepheid distances to supernova host galaxies without some slight modification. 
All of the Cepheid distance discussed so far are derived from using the reddening-free period-luminosity relation, the W(V,VI) Wesenheit function. A full description of this form of the PL relation and its history is given in the Appendix. We emphasize that the SHoES team also used a Wesenheit function, but choosing to form the reddening-free magnitude W(H,VI) by combining their H-band photometry with an appropriately scaled (V-I) color, whereas the Key Project used only W(V,VI). 
Accordingly, in this analysis we have calculated W(V,VI) PL relations for the SHoES galaxies and used these distances to compare with the TRGB distances available for 9 of them that have TRGB distances published so far. 
They are shown as red circles in the lower panel of Figures 1 and 2, and in the largest-distance portion of Figure 3. 

\begin{figure*}
\includegraphics[width=8.0cm,angle=-0]{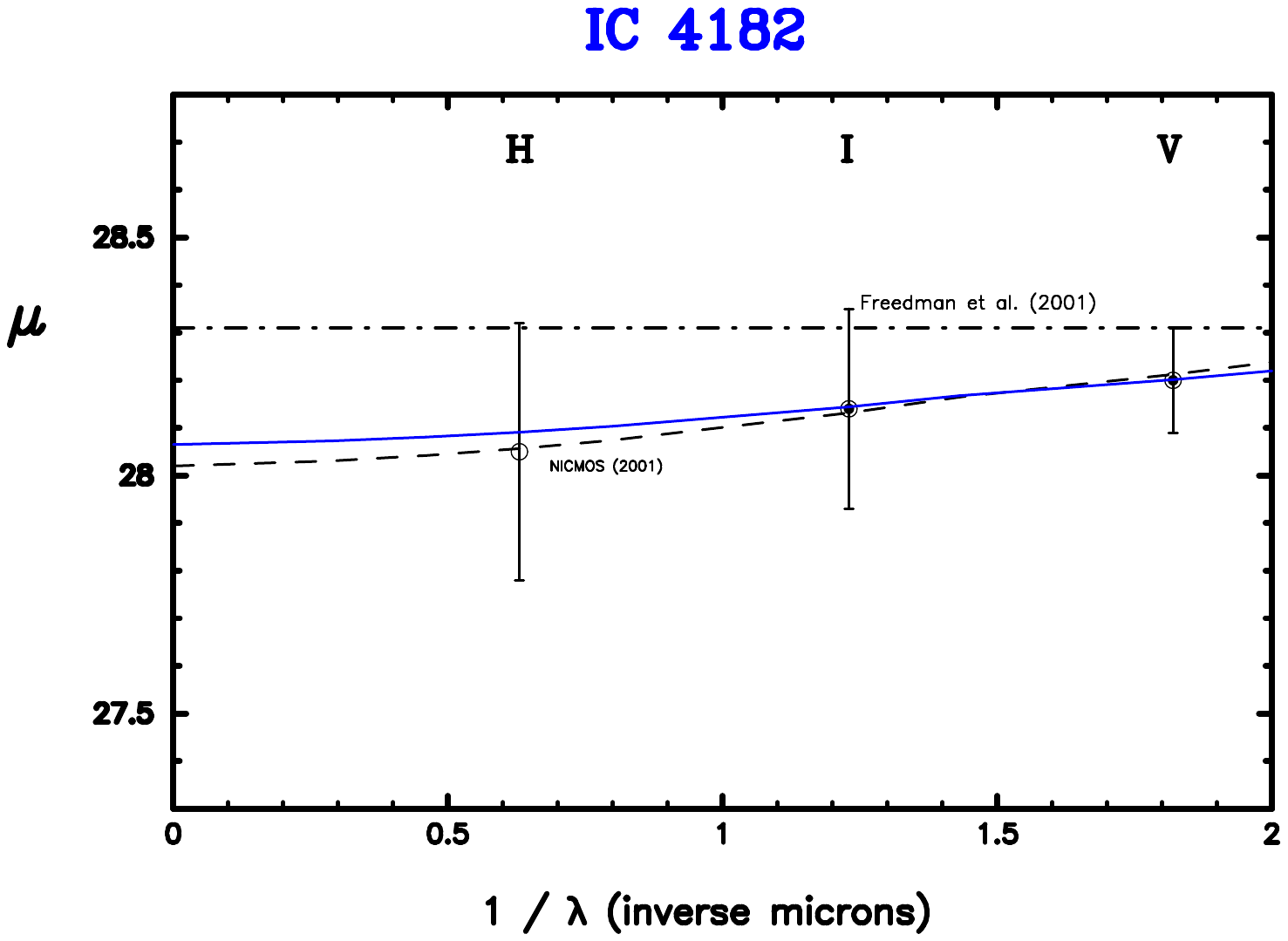}
\includegraphics[width=8.0cm,angle=-0]{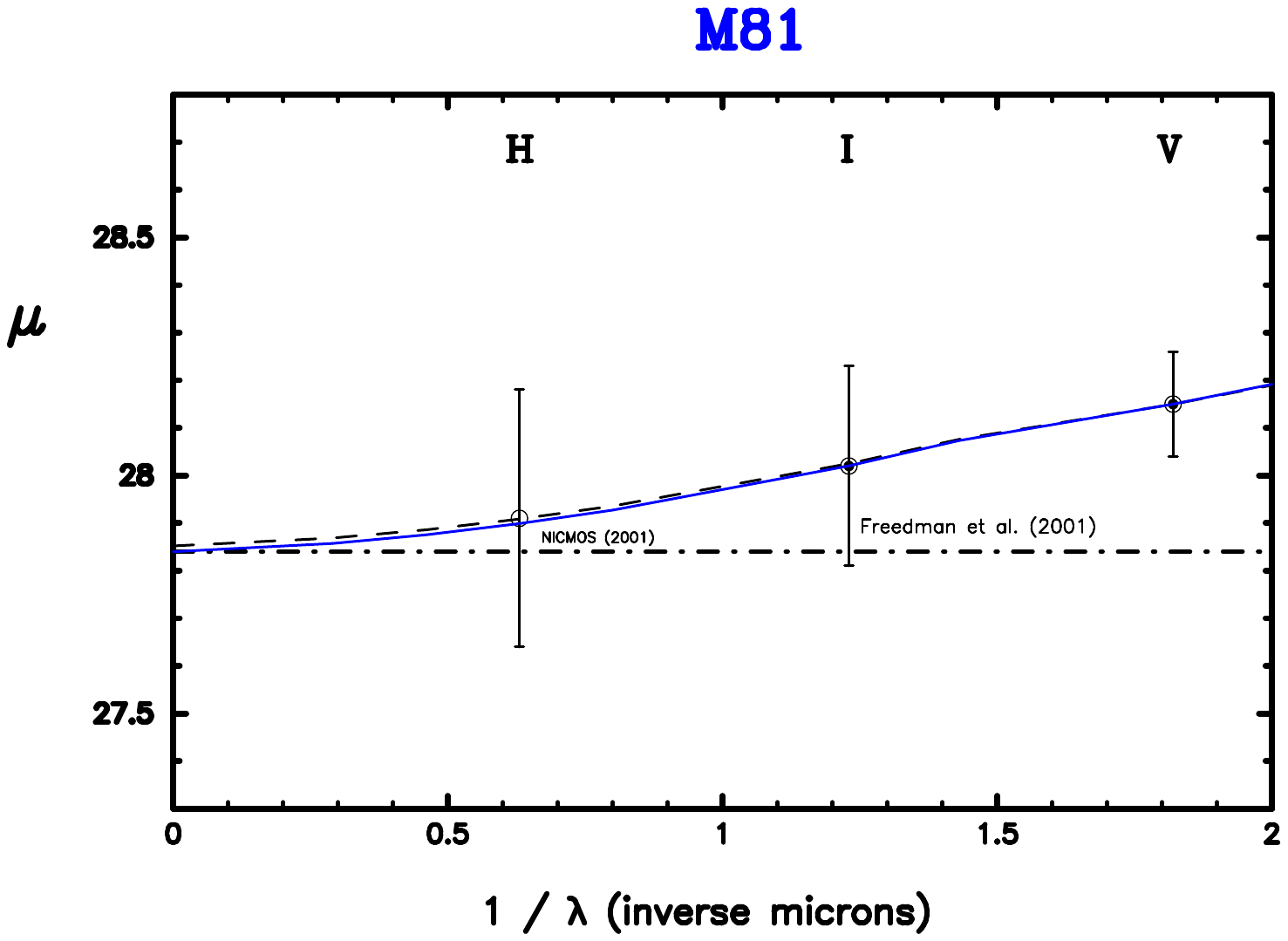}

\includegraphics[width=8.0cm,angle=-0]{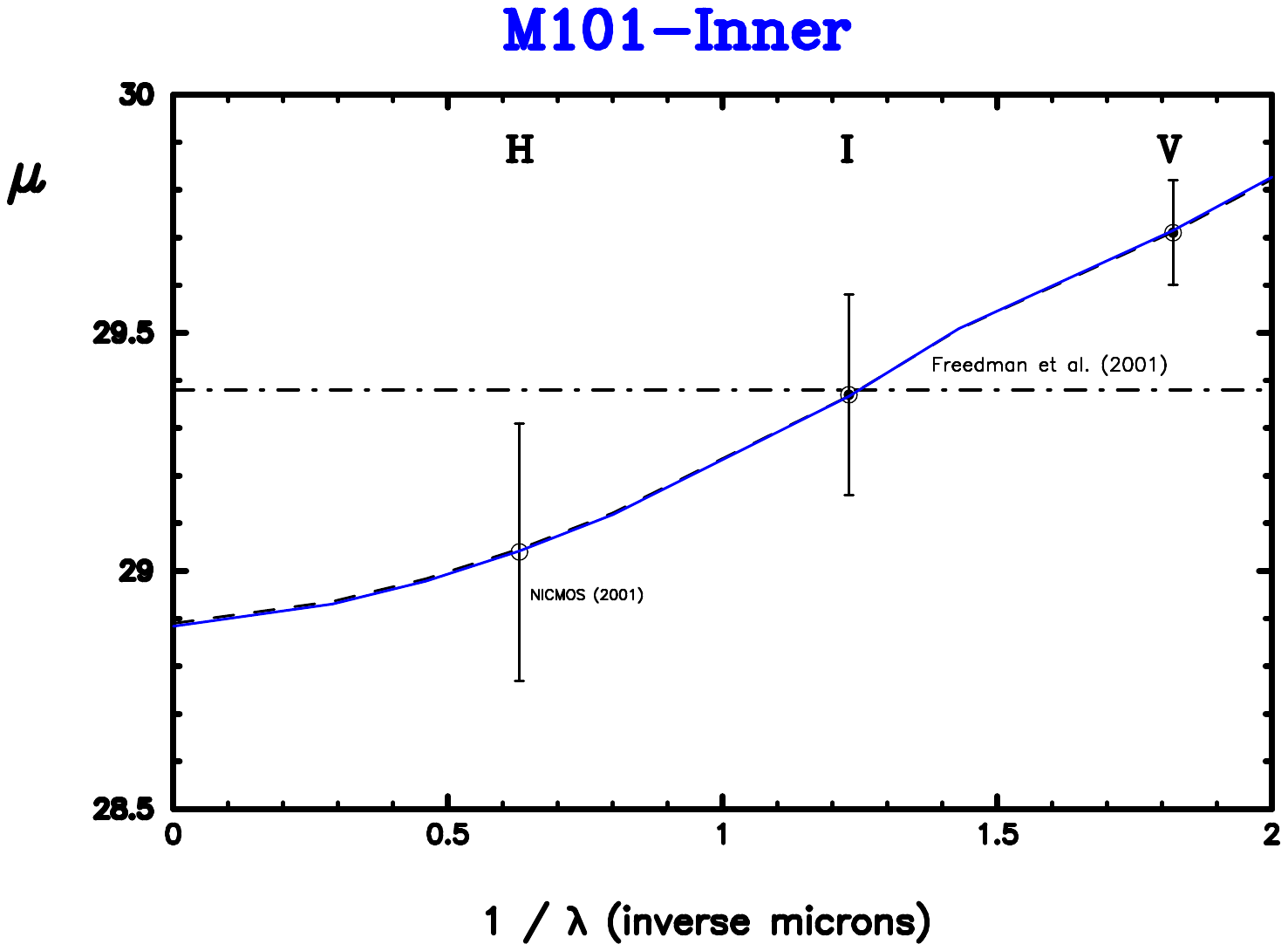}
\includegraphics[width=8.0cm,angle=-0]{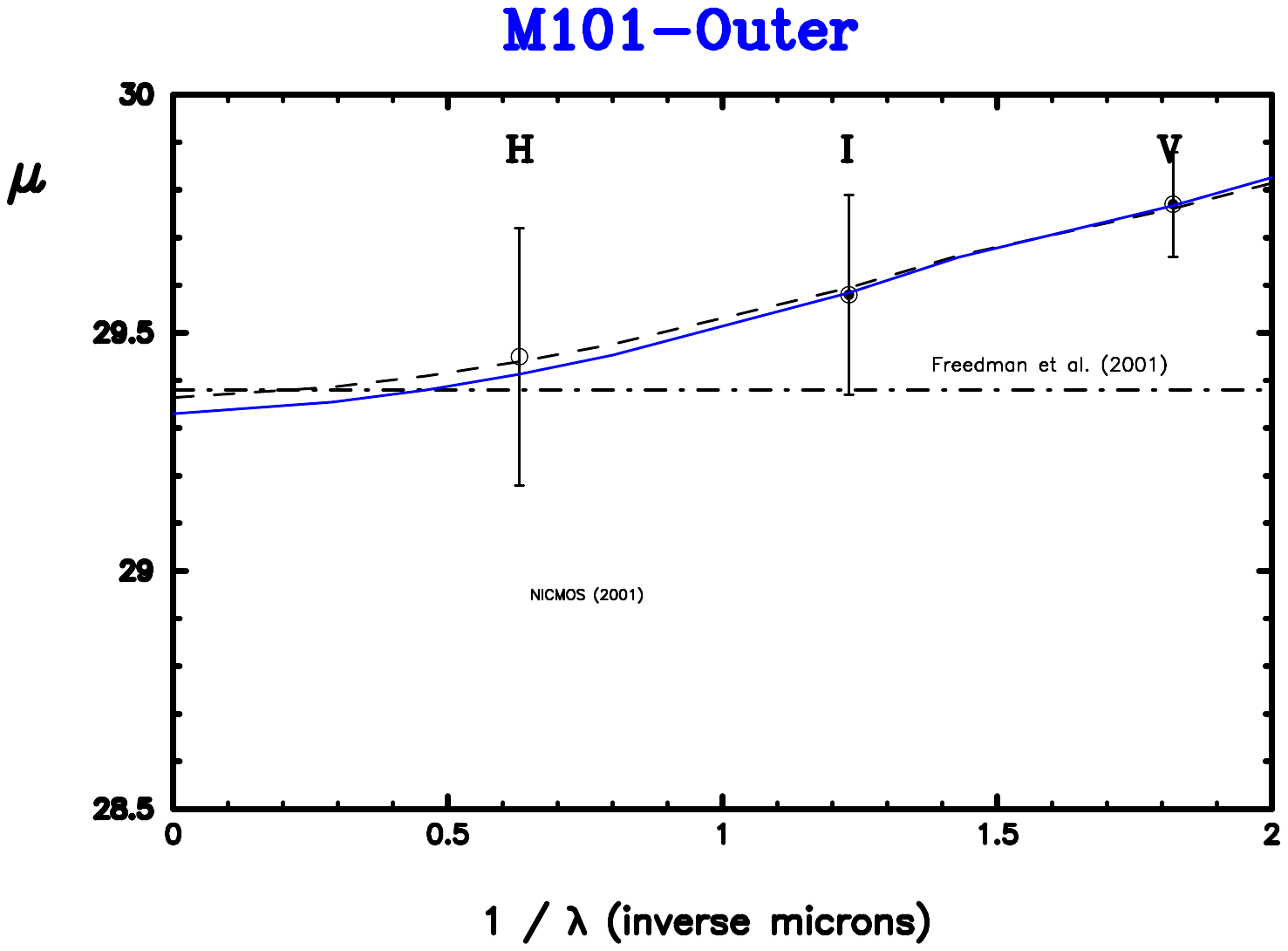}

\includegraphics[width=8.0cm,angle=-0]{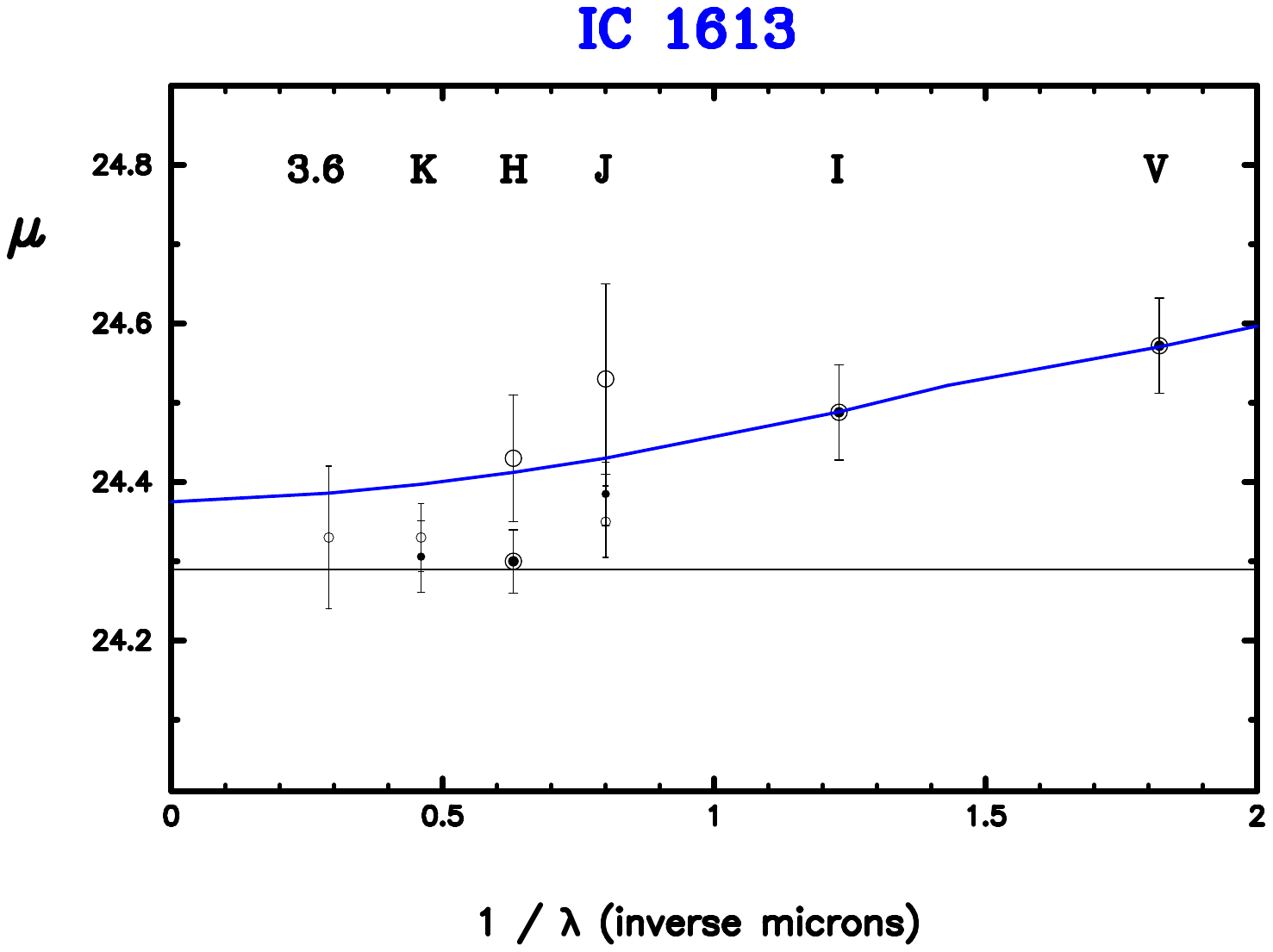}
\includegraphics[width=8.0cm,angle=-0]{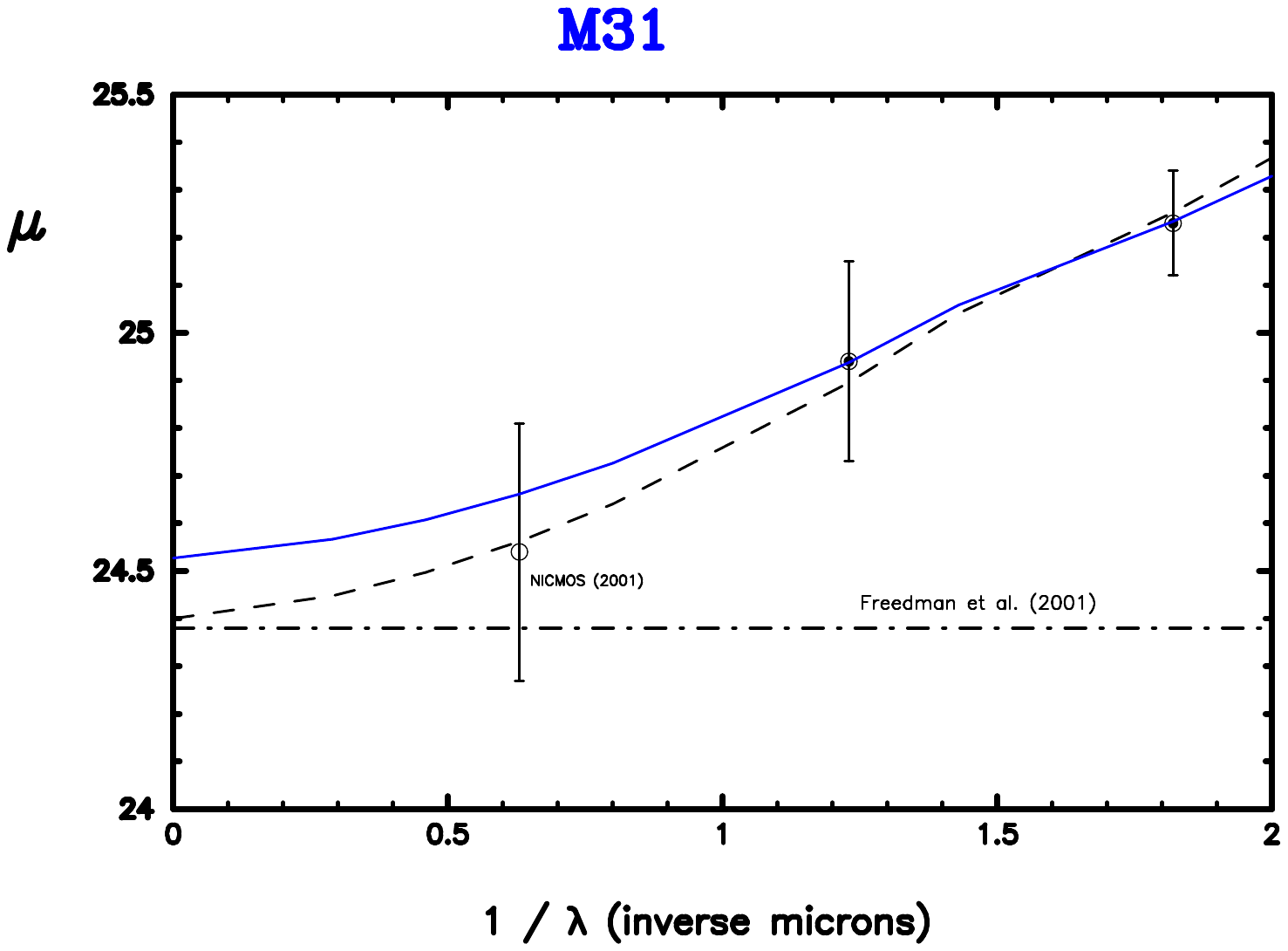}
\caption{\small Cepheid Extinction Curve Plots for Key Project galaxies  (Macri et al. 2001) following the installation of NICMOS on HST. Solid blue lines are fits to the VI data points as published earlier in Freedman et al. (2001). Dashed black lines show the three-point fit to the VIH data now including the NIR NICMOS observations. The horizontal  dash-dot lines show the distance modulus published in Freedman et al. (2001). However, for IC~1613, in particular, the large open circles are NICMOS data, while additional data points are taken from Scowcroft et al. (2017). The solid horizontal line is the multi-wavelength true distance modulus fit, as given in Scowcroft et al.. }
\end{figure*}

\begin{figure*}
\includegraphics[width=8.0cm,angle=-0]{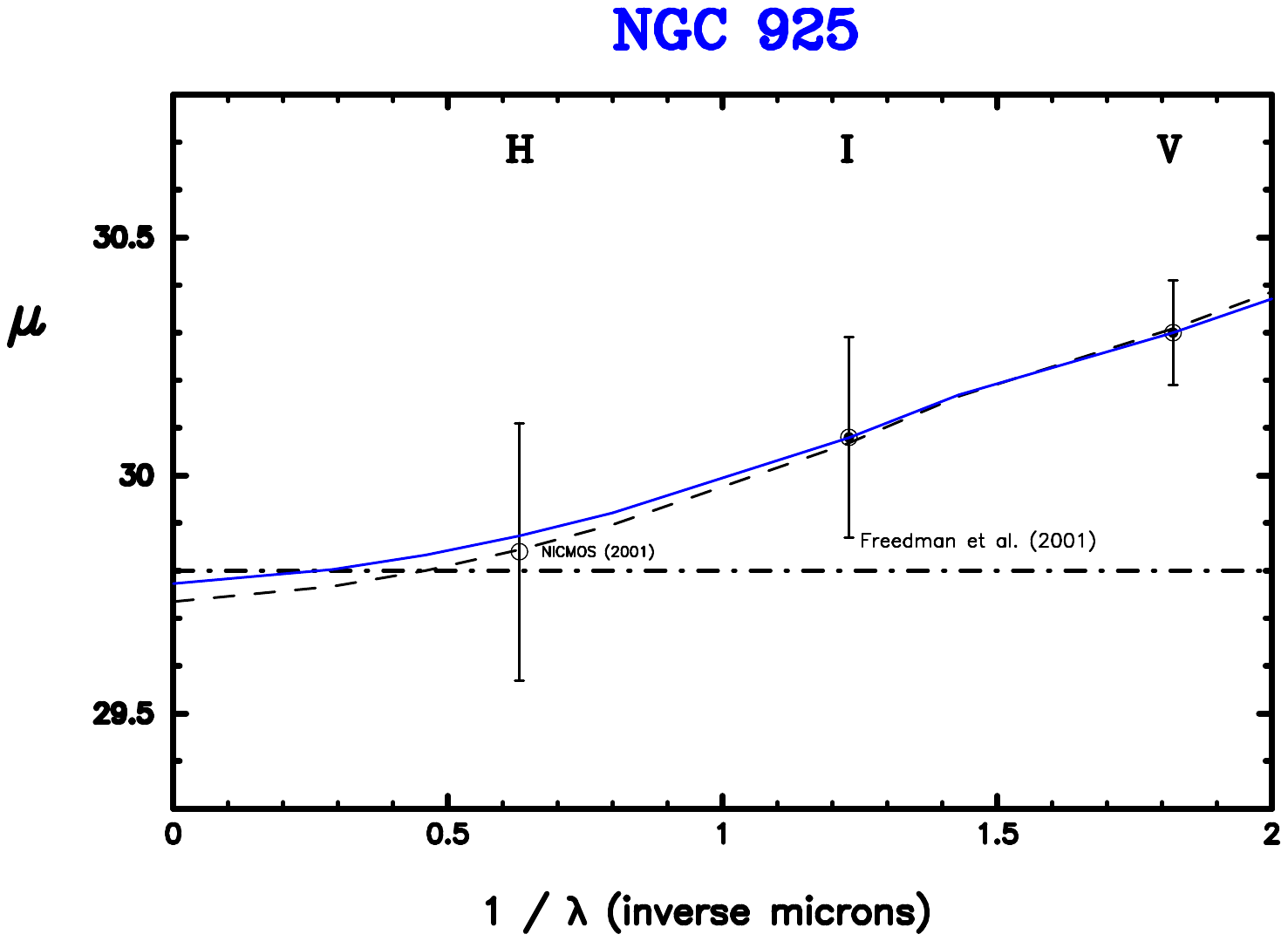}
\includegraphics[width=8.0cm,angle=-0]{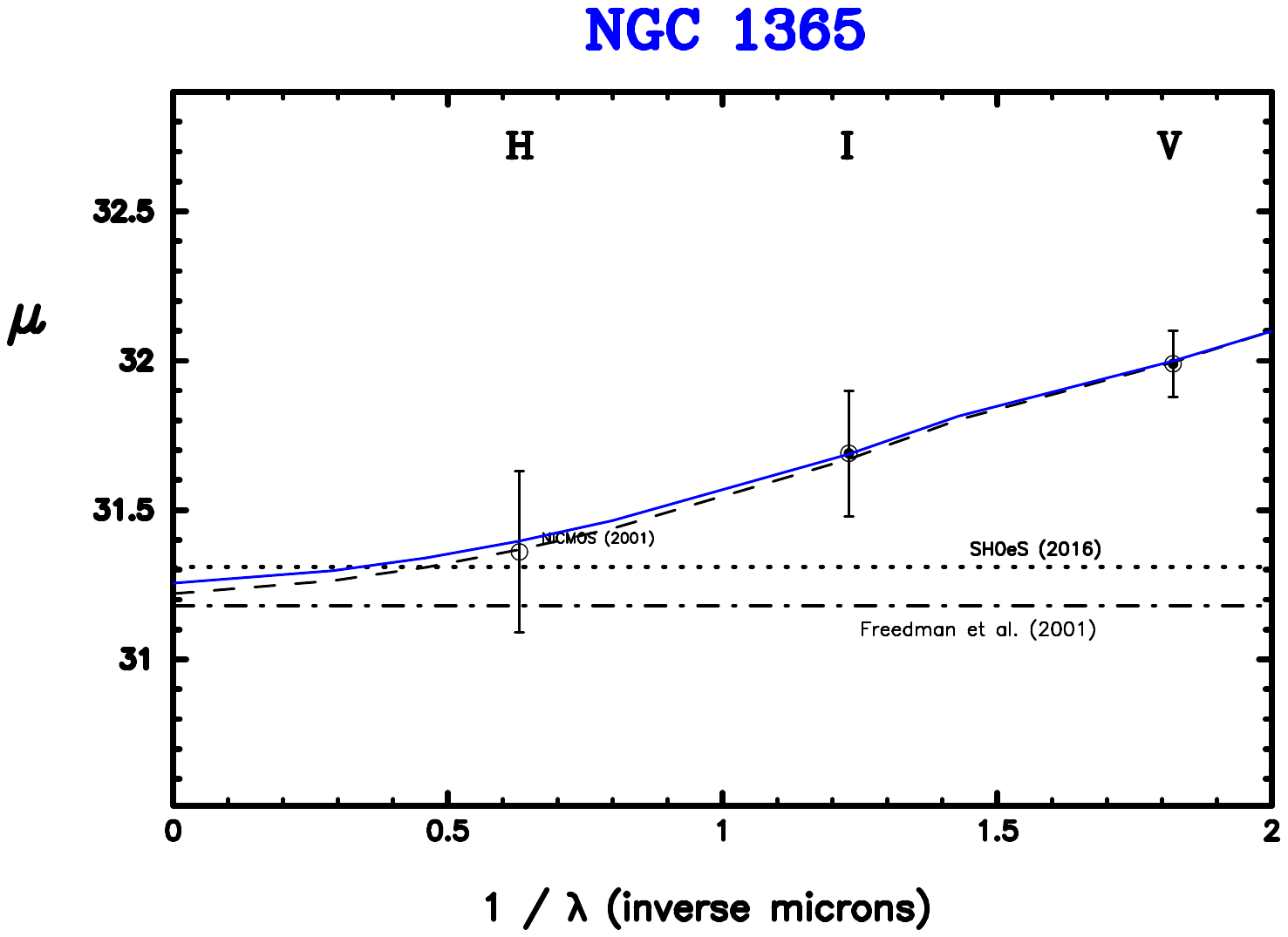}

\includegraphics[width=8.0cm,angle=-0]{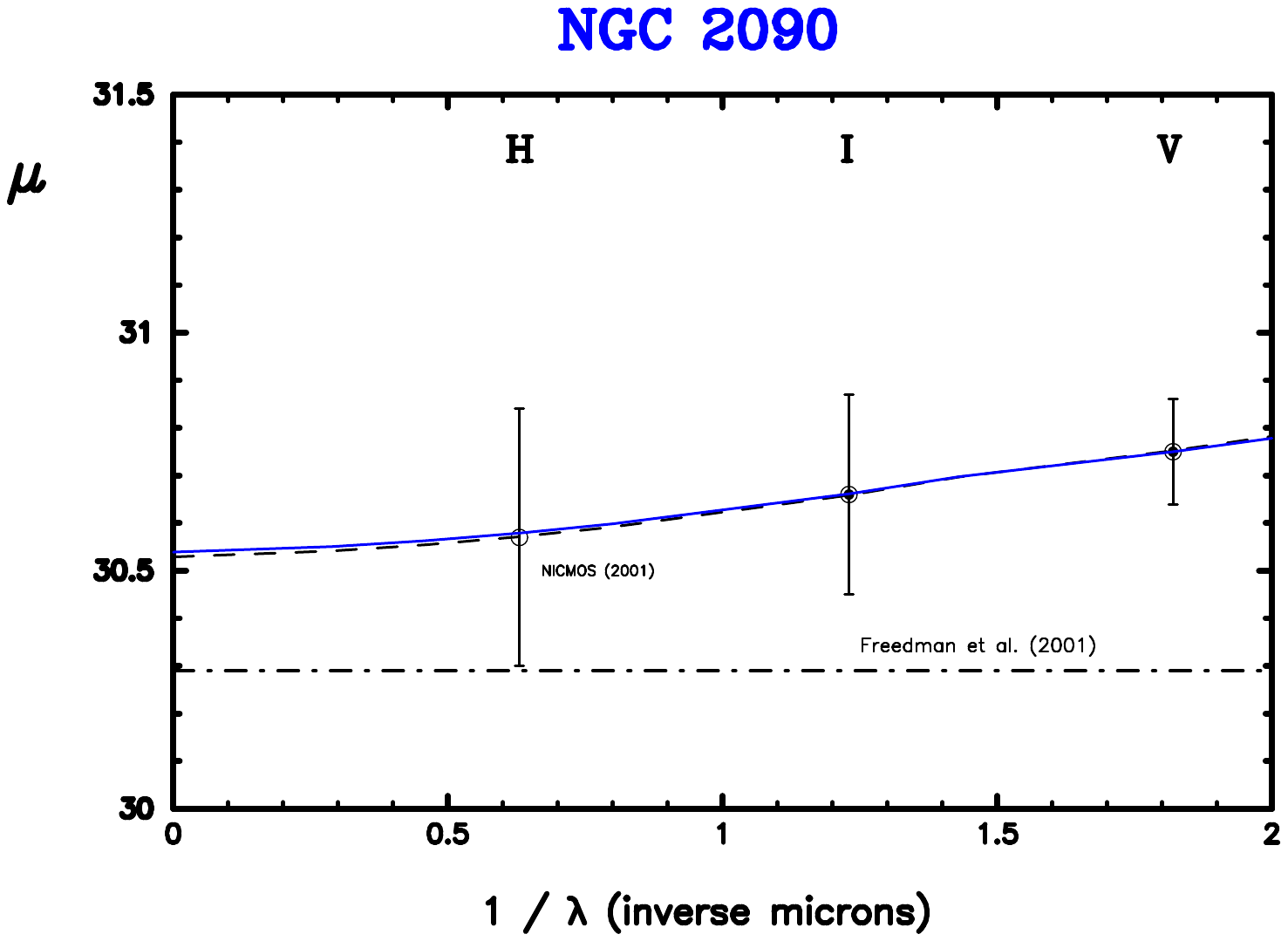}
\includegraphics[width=8.0cm,angle=-0]{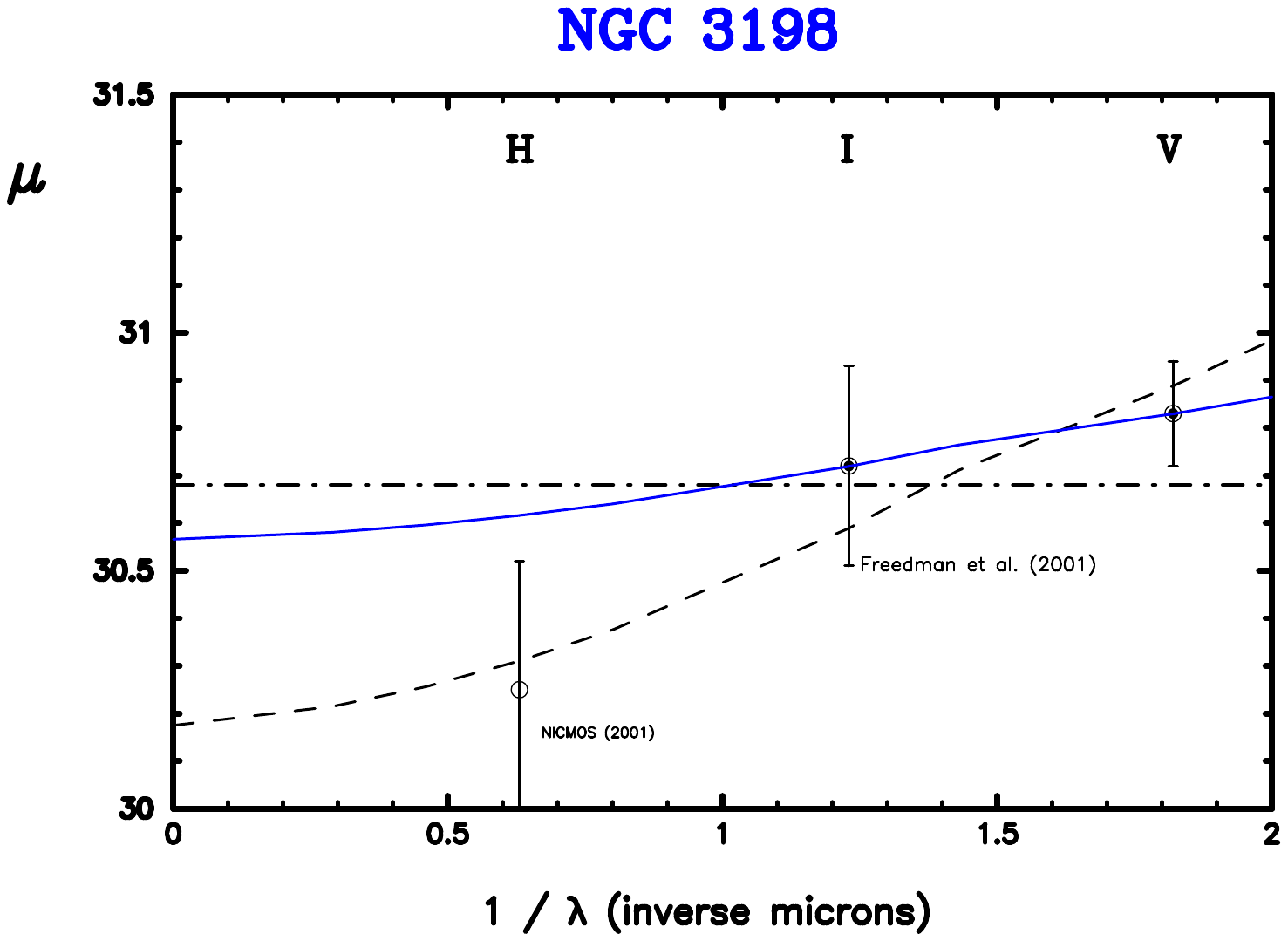}

\includegraphics[width=8.0cm,angle=-0]{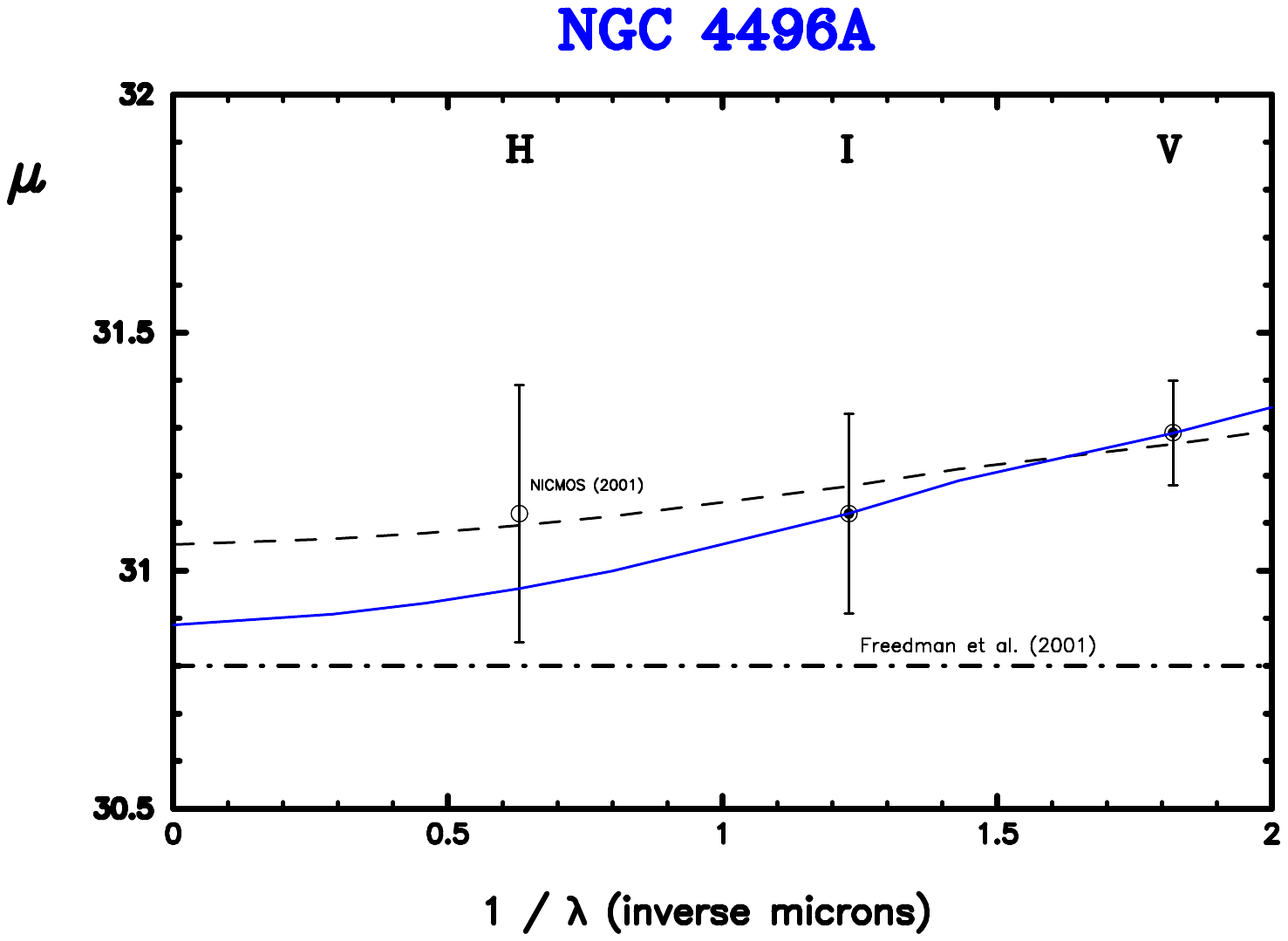}
\includegraphics[width=8.0cm,angle=-0]{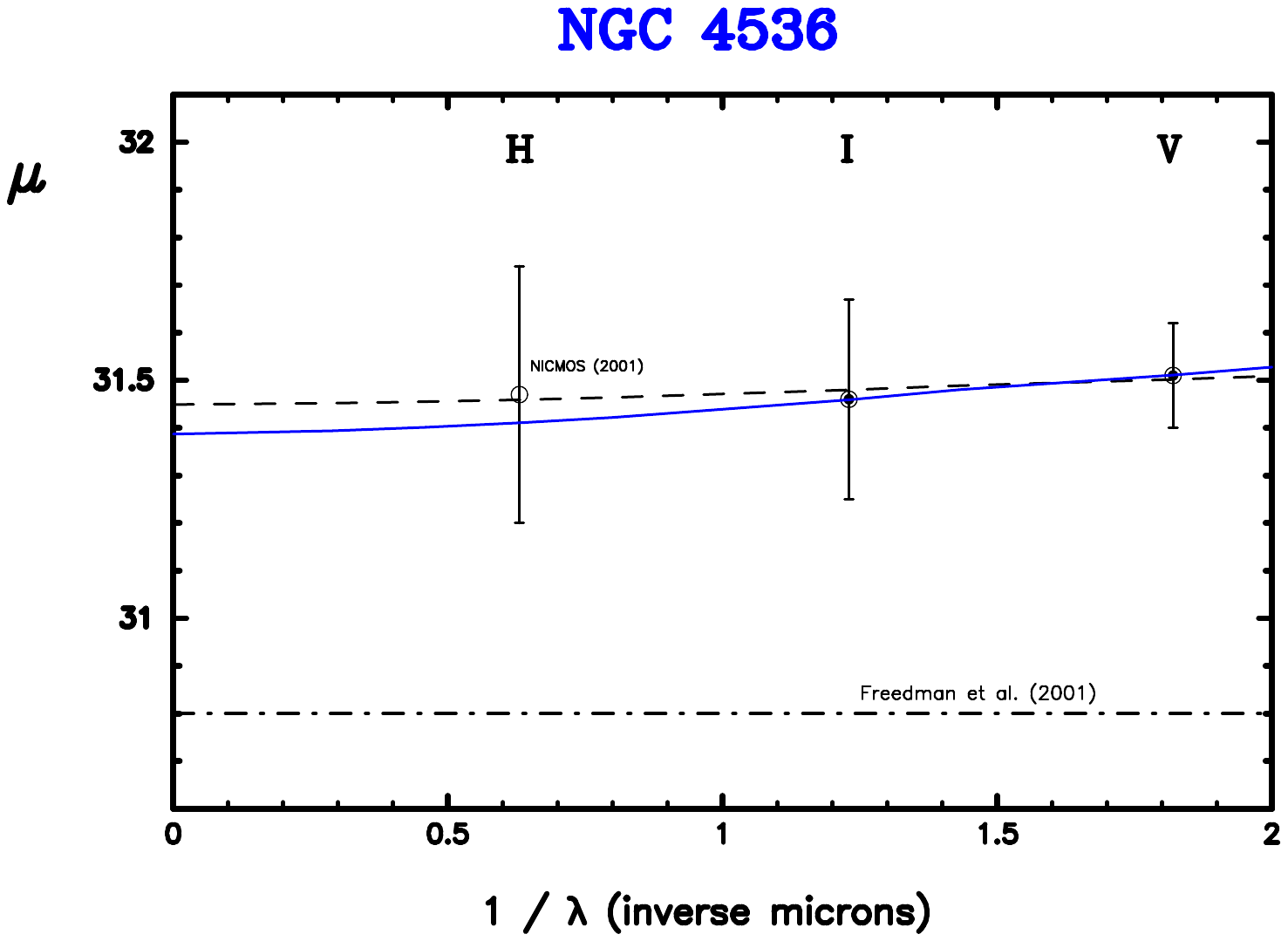}

\caption{\small Same as Figure 4}
\end{figure*}

The multi-wavelength fits of the Macri et al. (2001) data to a Milky Way extinction curve are shown in Figures 4 and 5. 
The SHoES data for Cepheids in the zero-point calibrating galaxy NGC~4258 are shown in Figure 6, where it is clear that the Cepheid distances derived from a variety of wavelength combinations have not yet convincingly converged, even for this relatively nearby, and very important zero-point calibrating galaxy. 
Figures 7 through 10 plot the complete set of SHoES (2016) apparent distance moduli fit here by a Milky Way extinction-curve using all three (VIH) bands (the dashed black line), and then only the VI data (the solid blue line). 
The Wesenheit W(H,VI) true distance modulus, as published in Riess et al. (2016), is shown by the red arrow to the far left in each plot. 
Values of the resulting W(V,VI) and W(H,VI) distance moduli are given in Table 2. 
A comparison of the VI and VIH extinction-curve fits are shown in Figure 9. 
There is no indication of any trend with distance modulus. 
The observed scatter $\pm$$ 0.16$ mag, which, if it is equally shared between the two methods, would suggest that the two Cepheid distance estimates are each good to $\pm$ 0.11~mag, or $\pm$6\% in distance.

The addition of the SHoES galaxies to Figures 1 and 2 has no obvious affect on the main conclusion of this paper, that there is little or no metallicity effect on the Cepheid distance based on W(V,VI) PL relations. The only deviant point belongs to the galaxy NGC~4424, which has very low statistical weight overall given that only 9 Cepheids were found in it in the SHoES survey. The main takeaway from Figure 2, however, is that the more distant SHoES galaxies have significantly larger scatter in their comparison with the TRGB distances ($\pm$0.152 mag for the distant sample, as opposed to $\pm$0.066 mag for the more nearby galaxies), and that there is an indication of an offset of about a tenth of a magnitude in the mean of the two samples.

\begin{figure*} 
\centering 
\includegraphics[width=14.0cm,angle=-0]{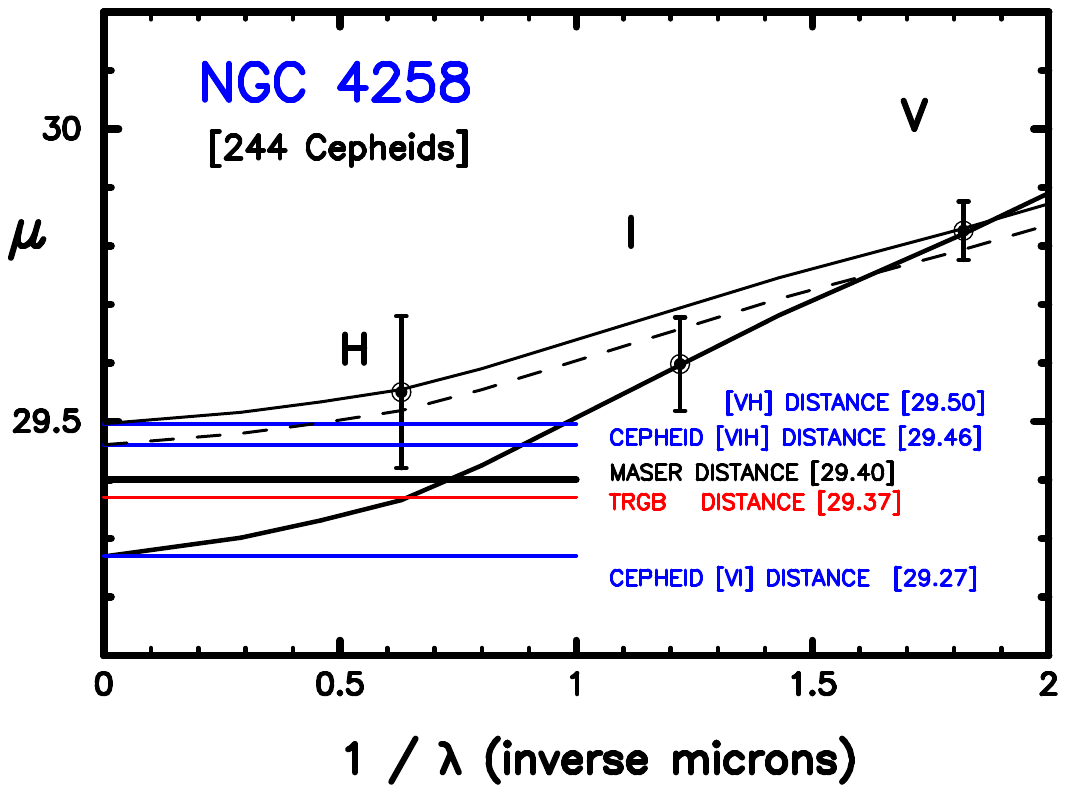}
\caption{\small Extinction Curve Fits to VI (thick black line), VH (thin black line)and VIH (broken black line) data for Cepheids in the maser galaxy NGC~4258. The corresponding Cepheid true distance moduli, three in number, are given in blue. The TRGB true distance modulus is given in red. And the geometric/maser distance modulus to NGC~4258 is given in black. The total spread in true distance moduli is more than 0.2~mag.}
\end{figure*}

\vfill\eject

\begin{figure*} 
\includegraphics[width=8.0cm,angle=-0]{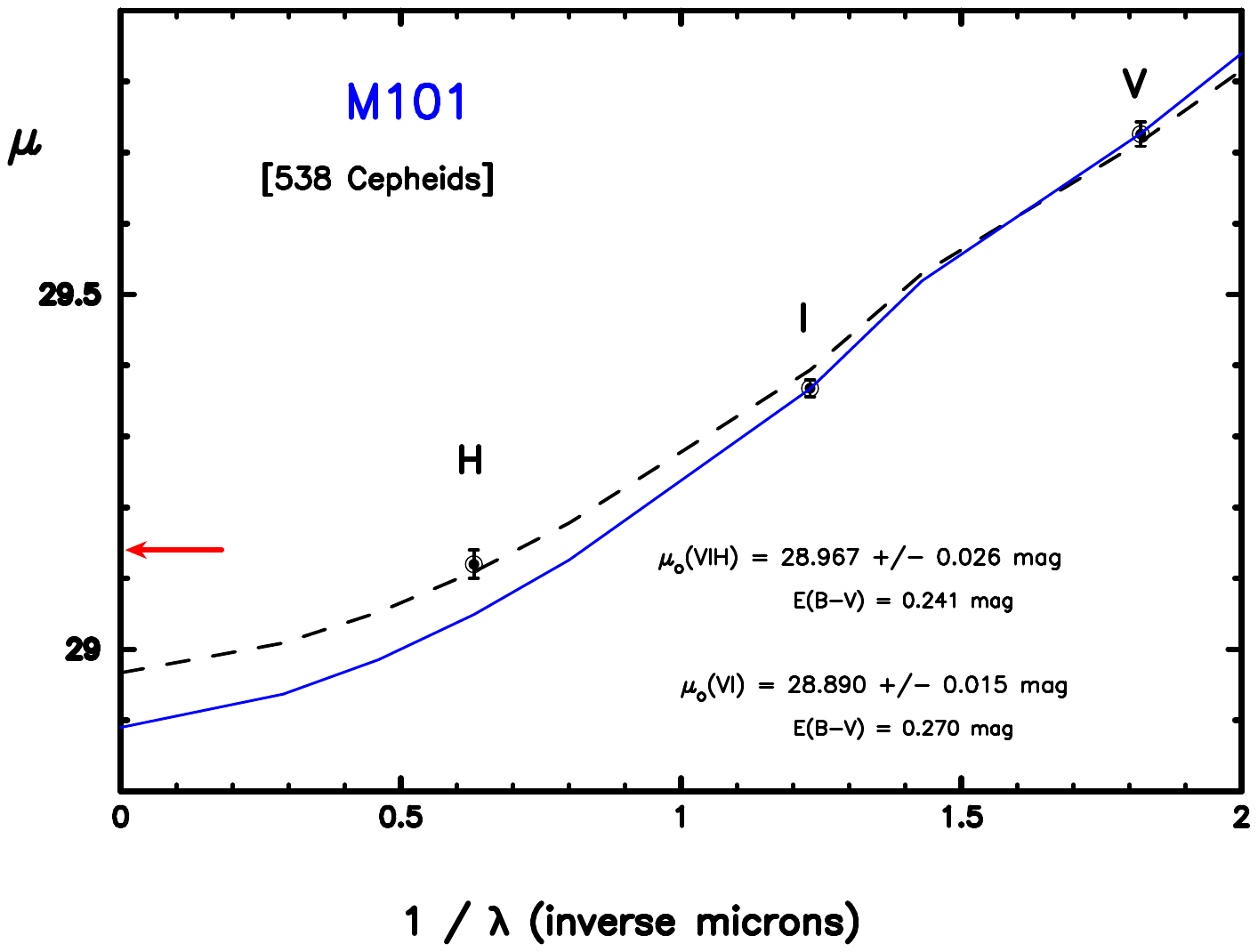}
\includegraphics[width=8.0cm,angle=-0]{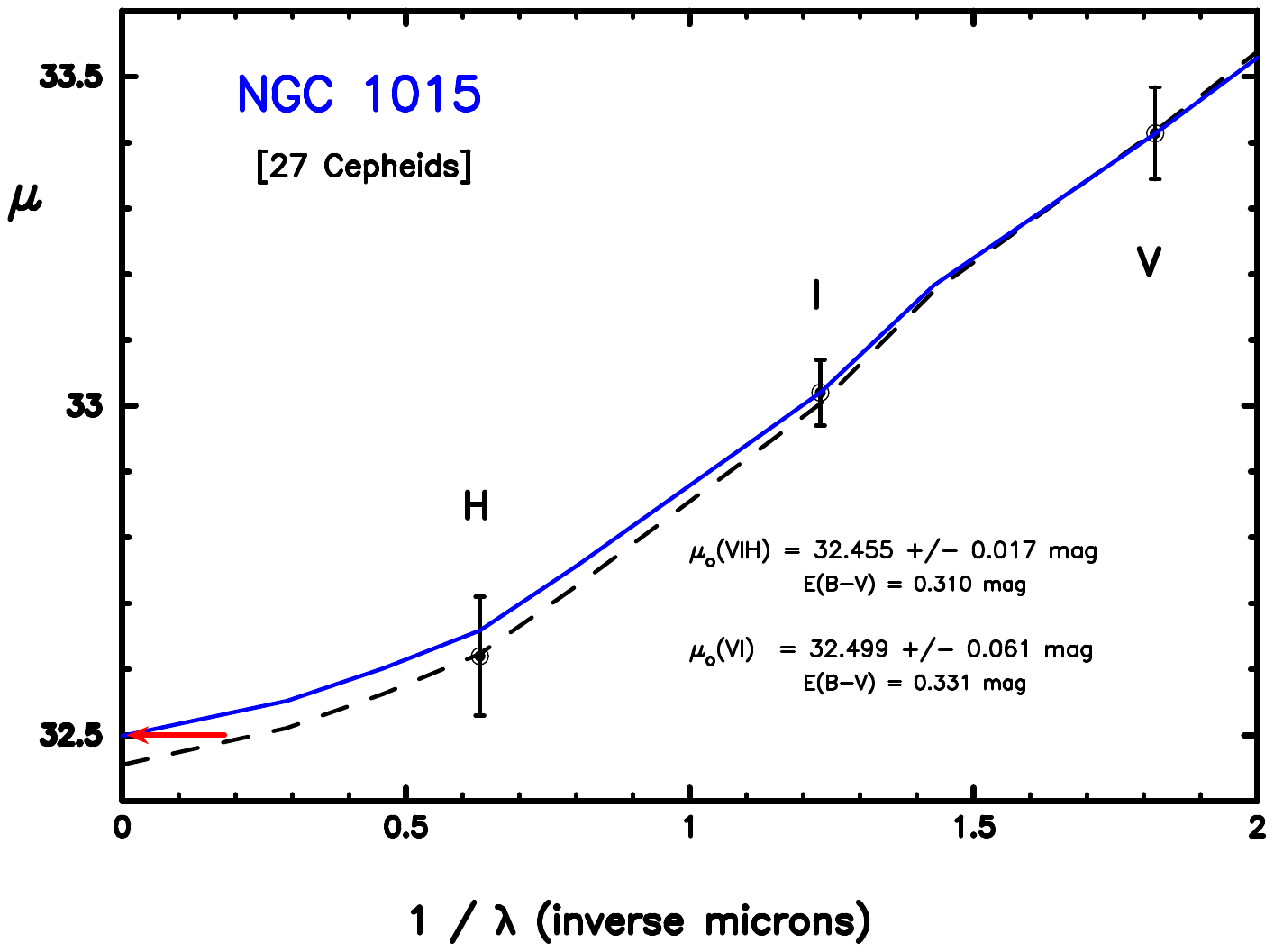}

\includegraphics[width=8.0cm,angle=-0]{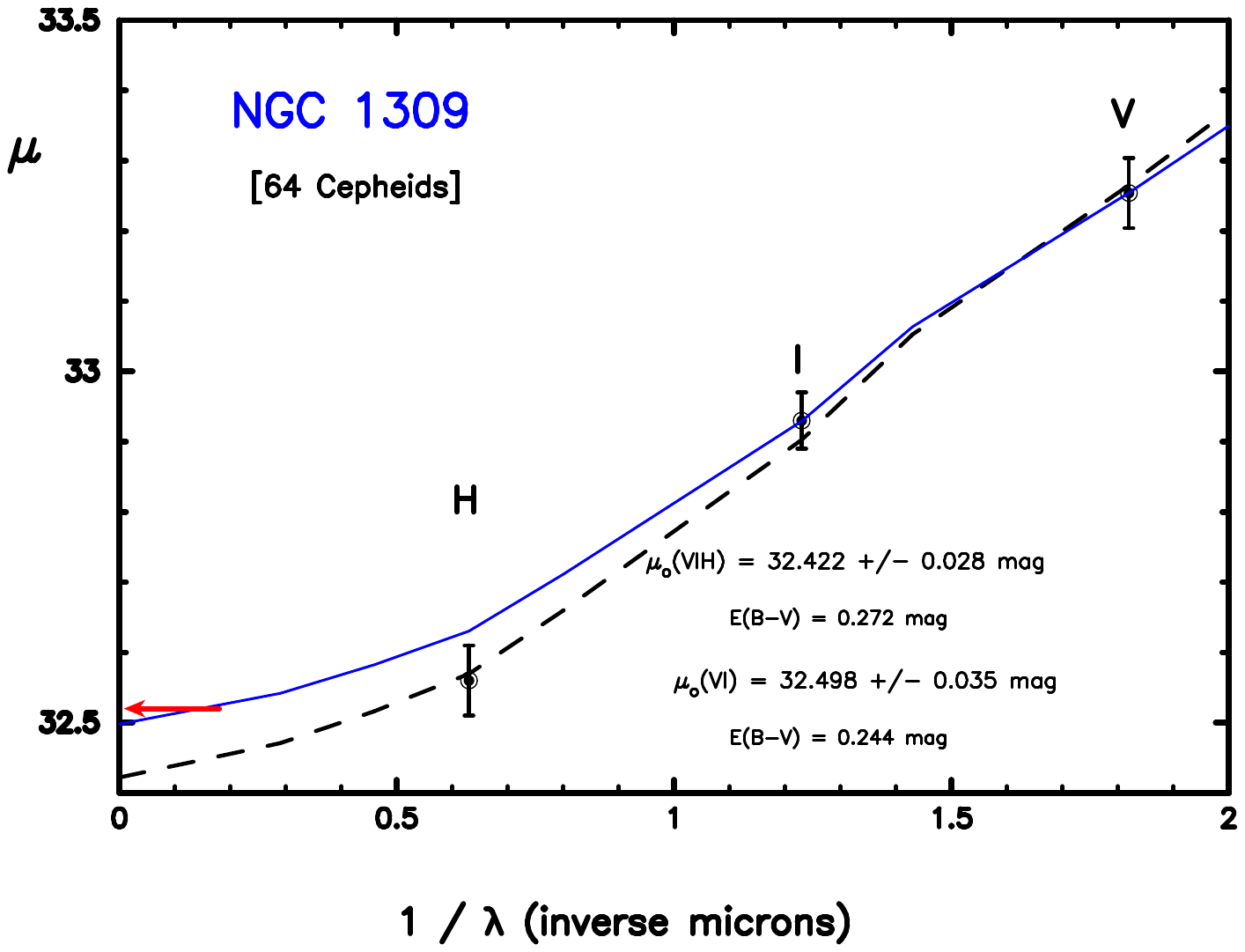}
\includegraphics[width=8.0cm,angle=-0]{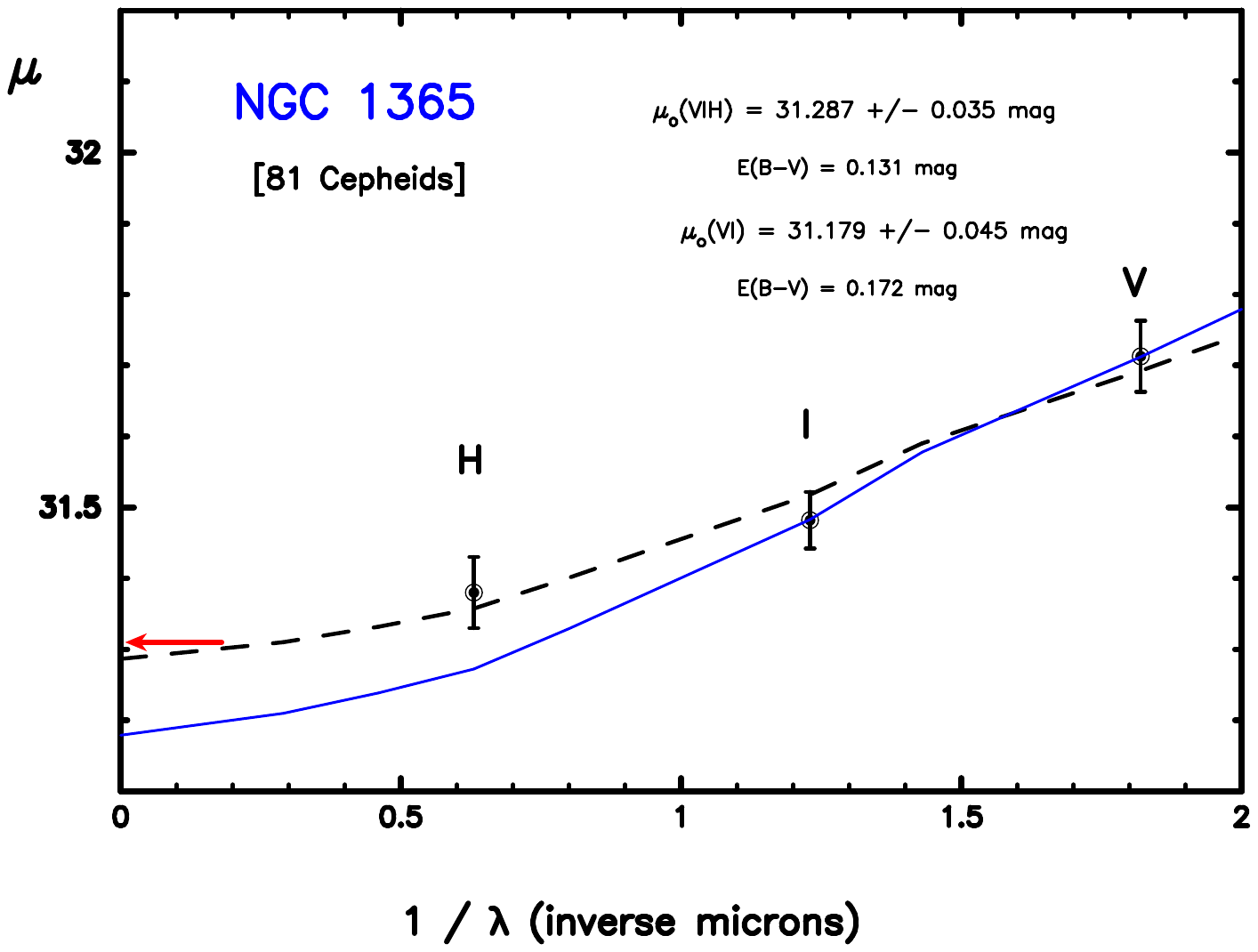}

\includegraphics[width=8.0cm,angle=-0]{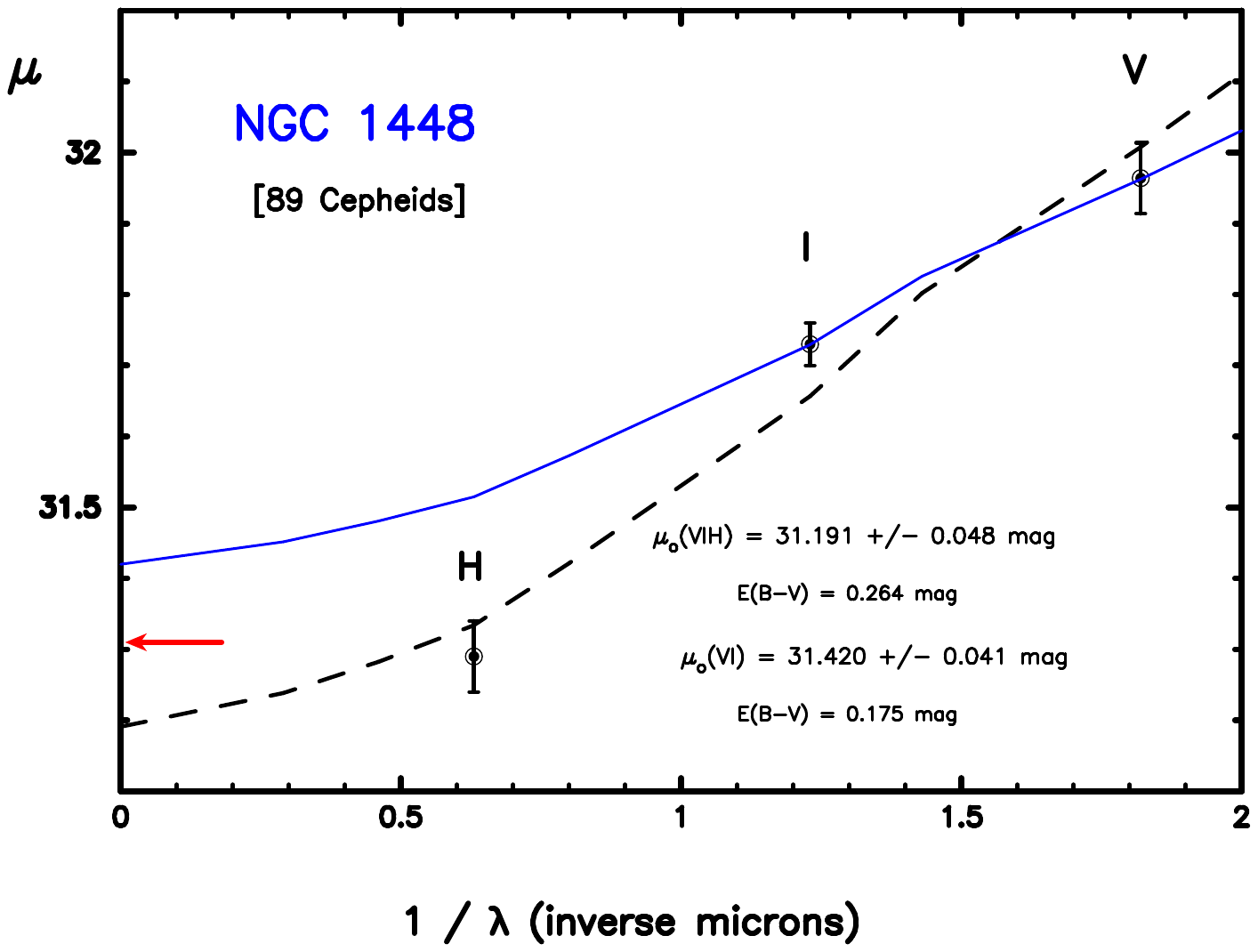}
\includegraphics[width=8.0cm,angle=-0]{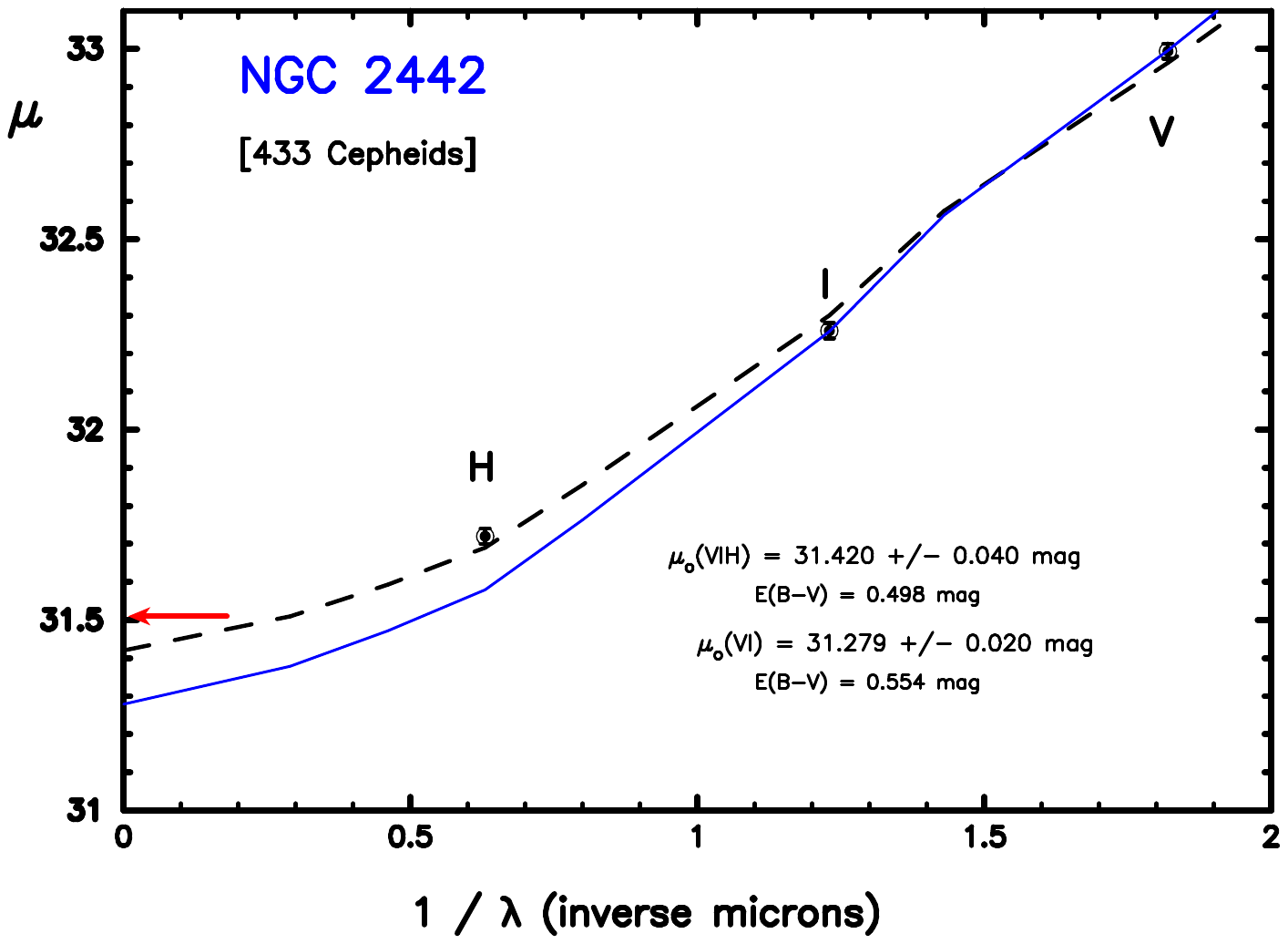}
\caption{\footnotesize Extinction Curve Fits to VI and VIH data for SHoES galaxies hosting Type Ia SNe in  Riess et al. (2016). The red arrow at the Y axis indicates the SHoES distance modulus. A fit to the VI data alone is shown by a solid blue line; while a fit to the combined VIH data is shown by a dashed black line. The corresponding true moduli and mean reddenings are explicitly given in the lower right corner of each plot, and given again in Table 2.}
\end{figure*}

\begin{figure*} 
\includegraphics[width=8.0cm,angle=-0]{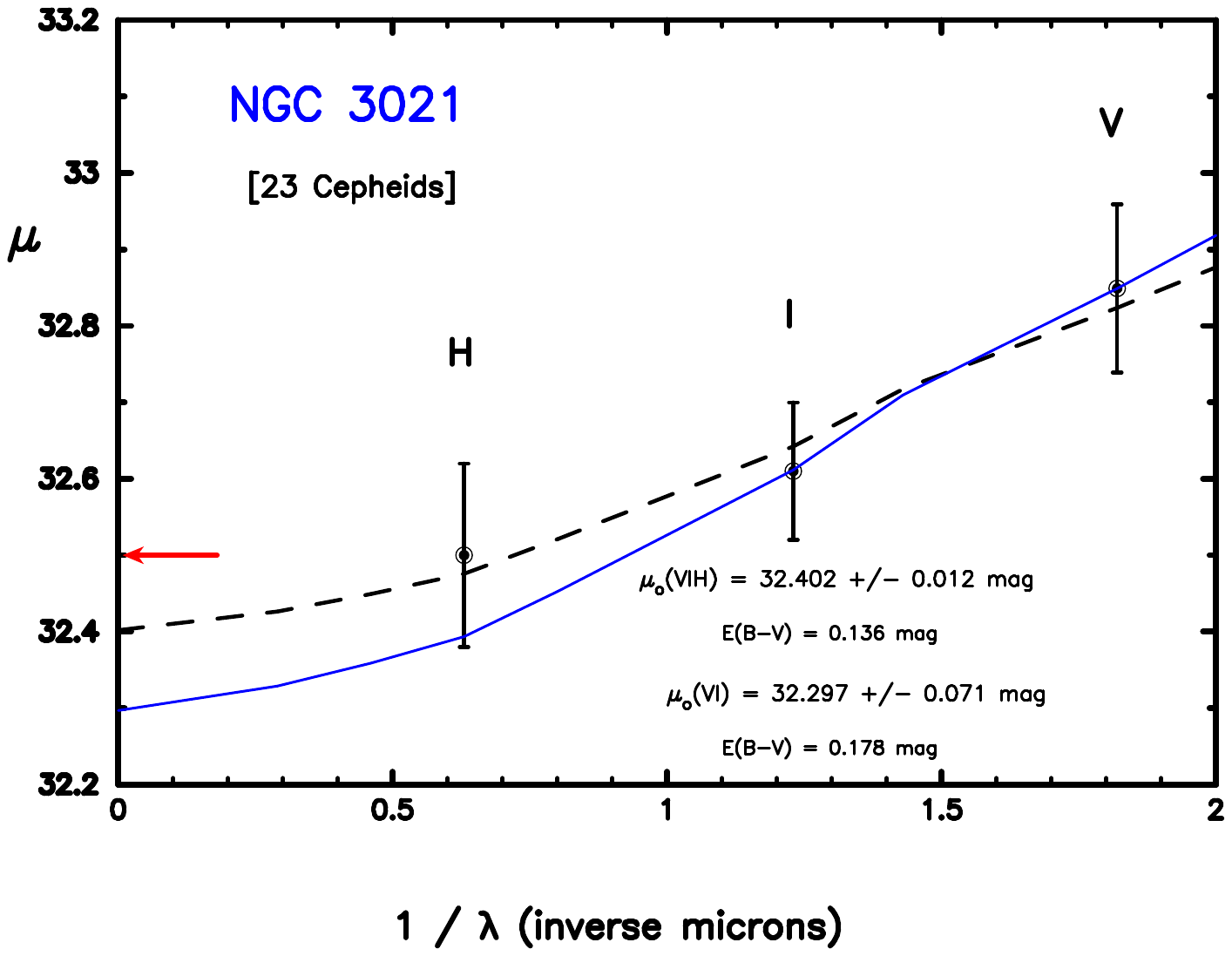}
\includegraphics[width=8.0cm,angle=-0]{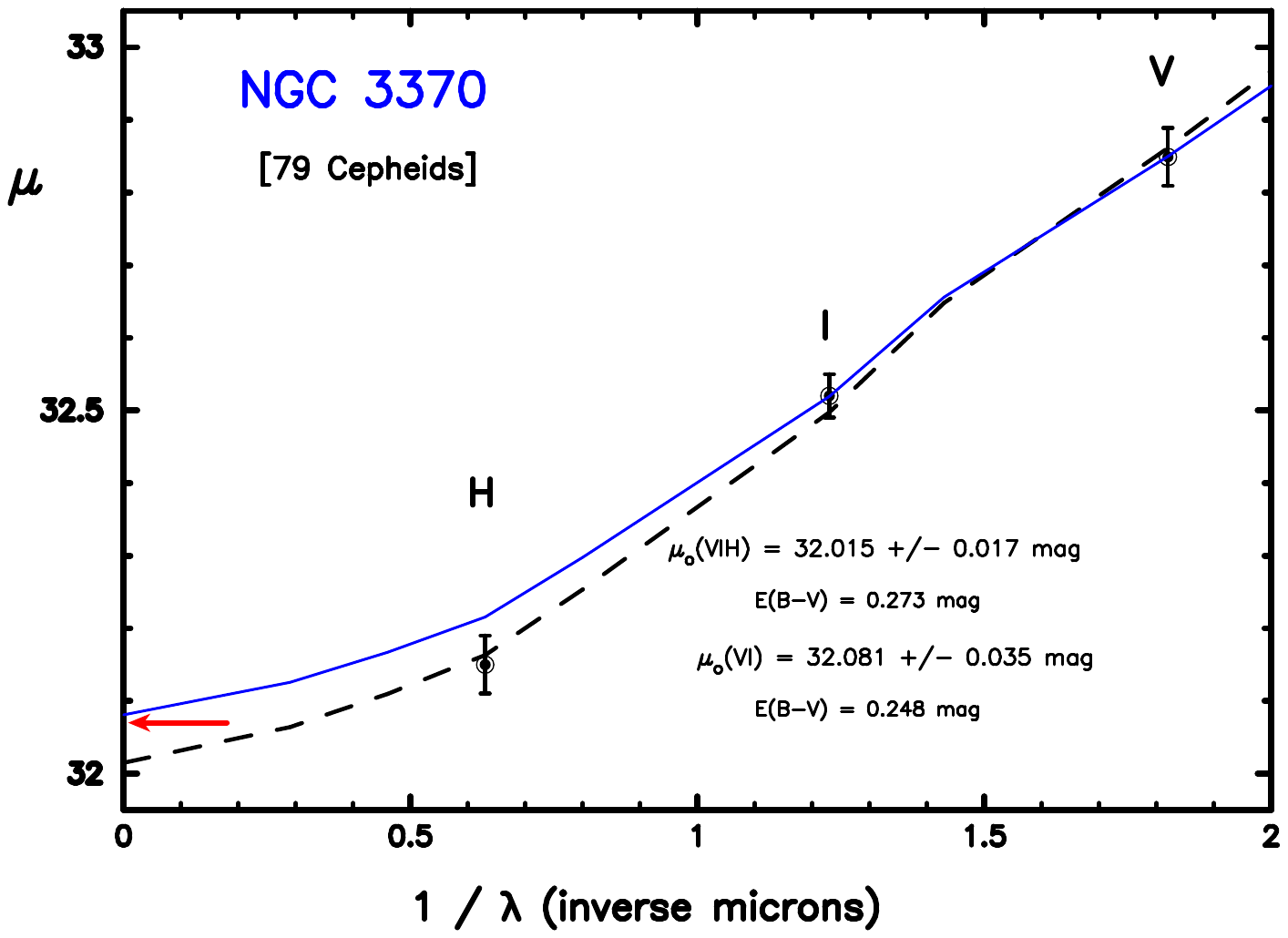}

\includegraphics[width=8.0cm,angle=-0]{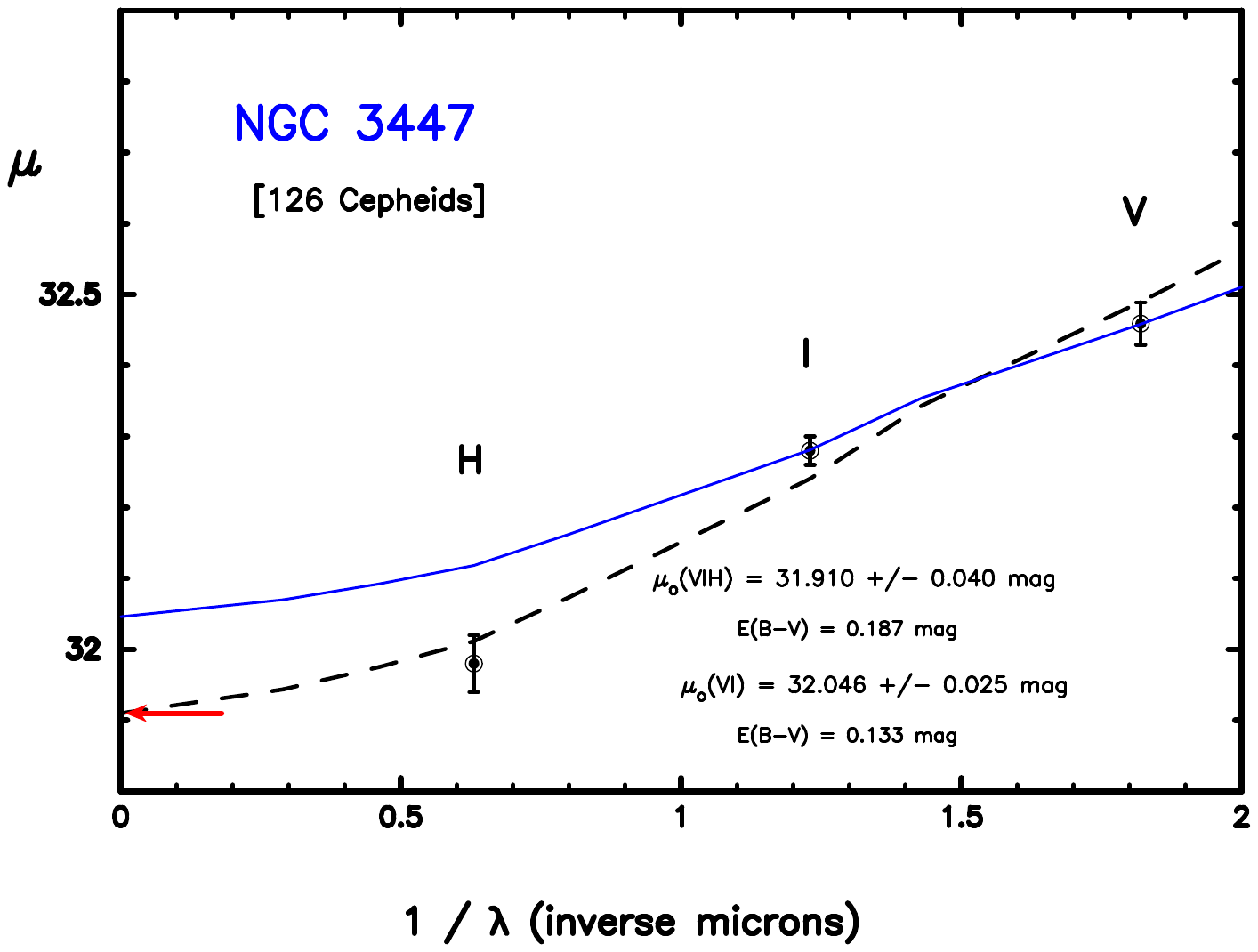}
\includegraphics[width=8.0cm,angle=-0]{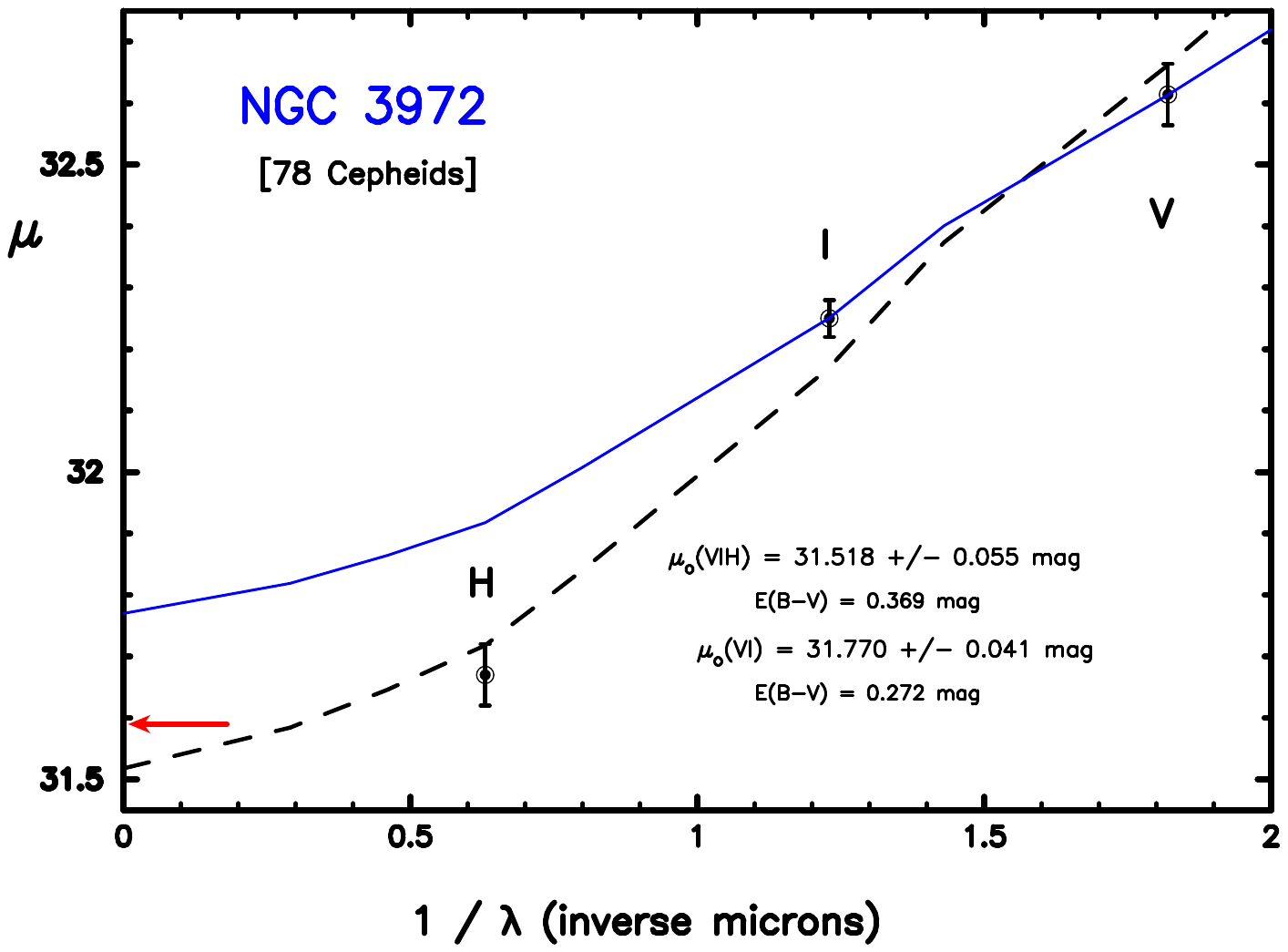}

\includegraphics[width=8.0cm,angle=-0]{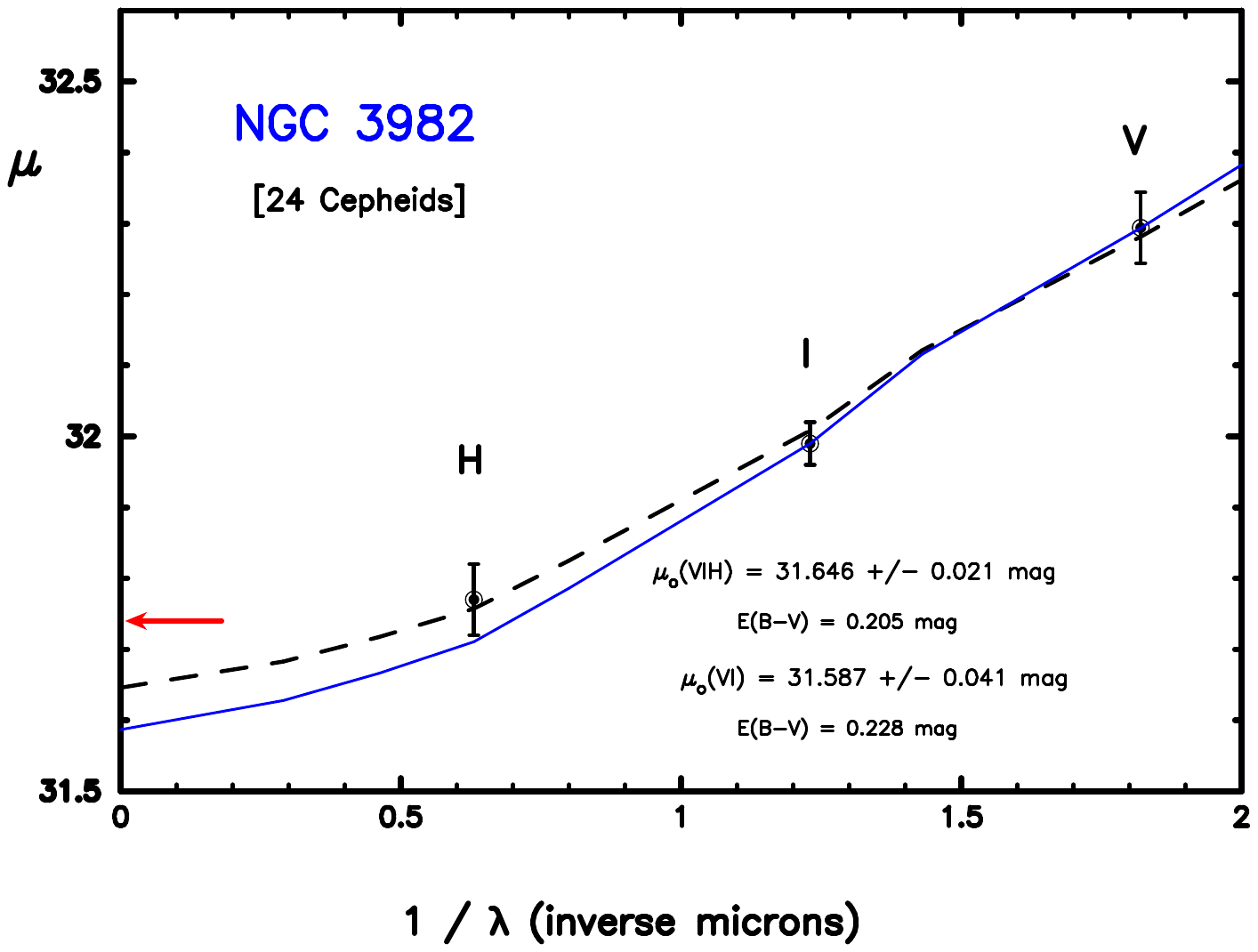}
\includegraphics[width=8.0cm,angle=-0]{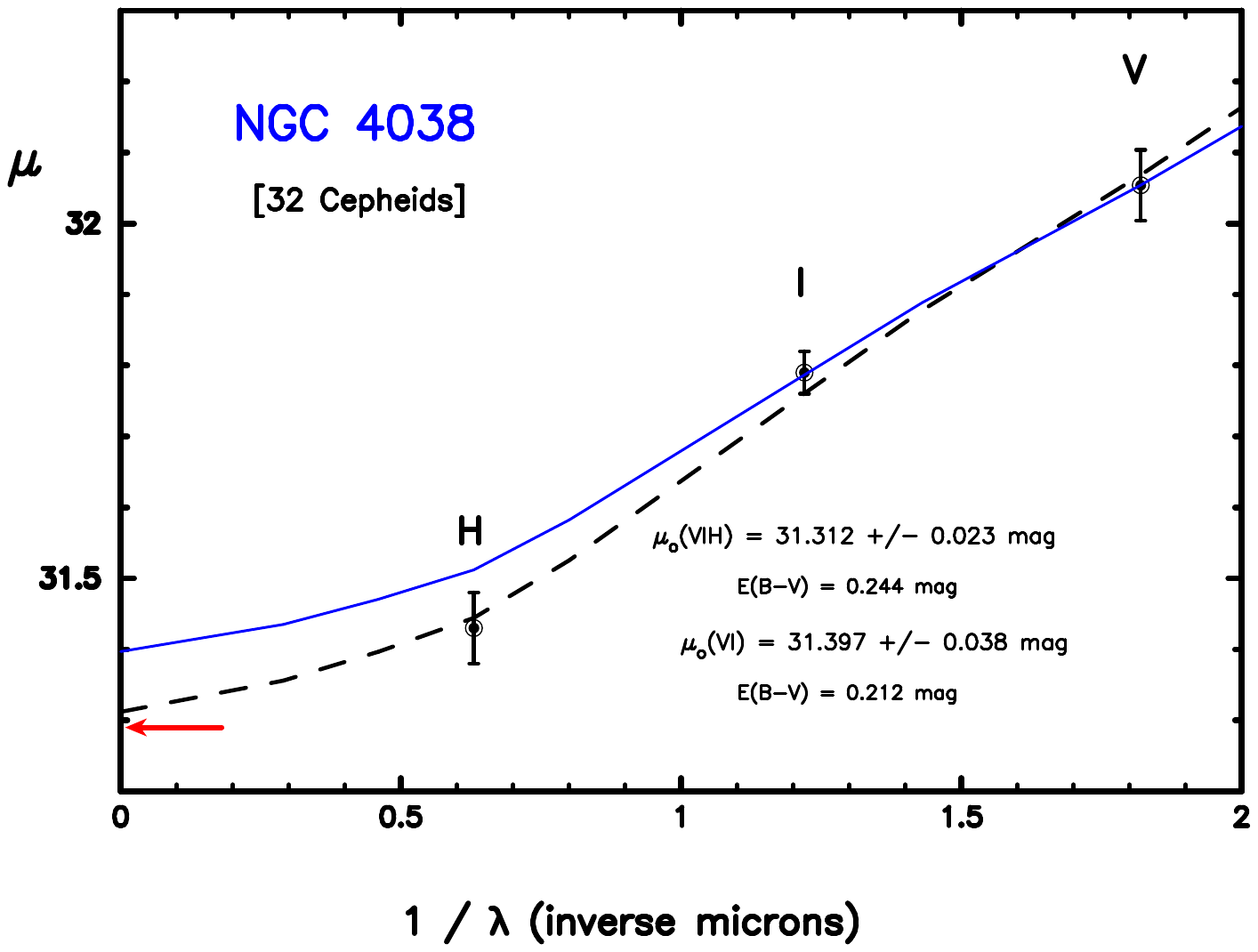}

\caption{Same as for Figure 7.}
\end{figure*}

\begin{figure*}
\includegraphics[width=8.0cm,angle=-0]{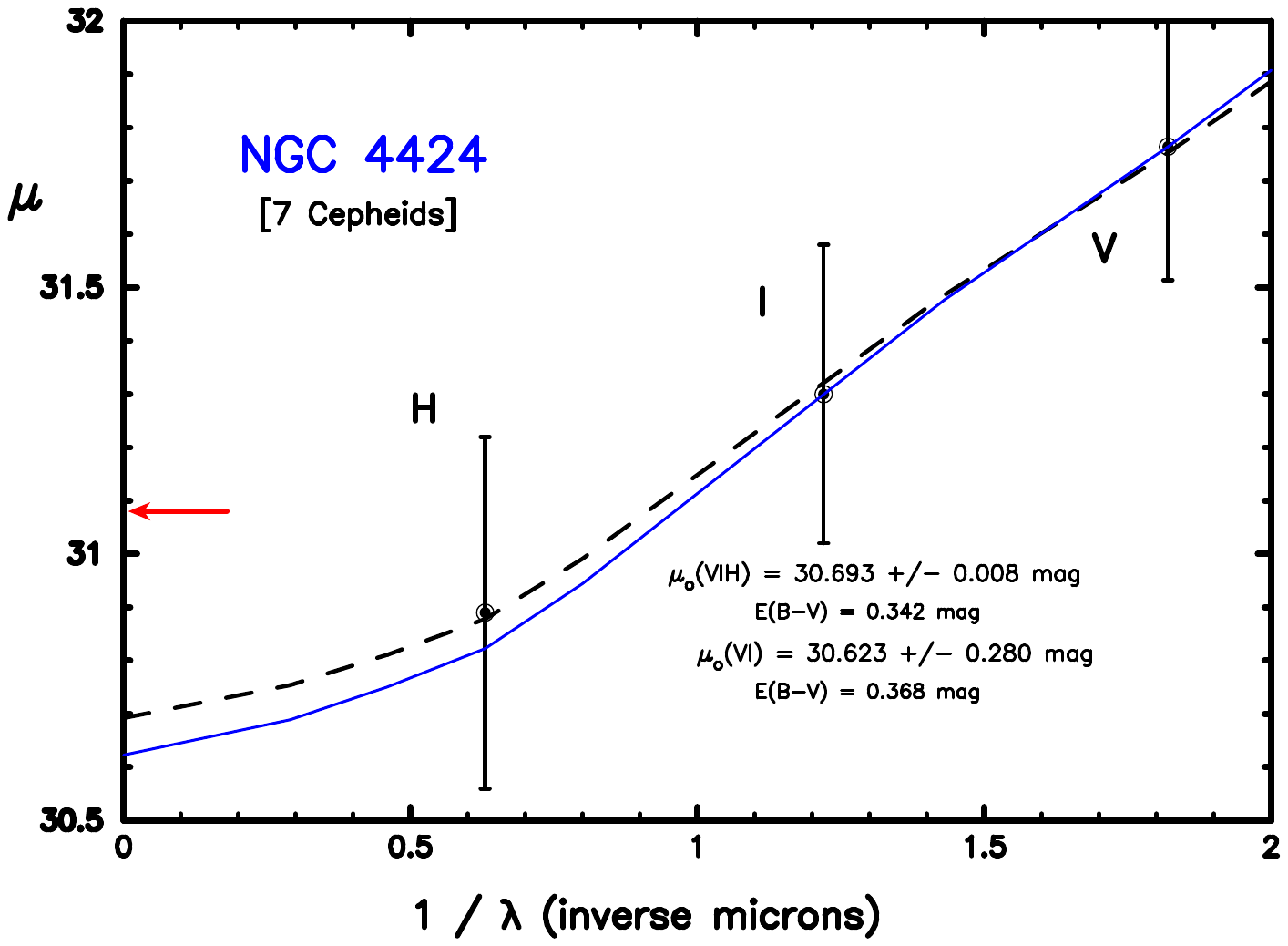}
\includegraphics[width=8.0cm,angle=-0]{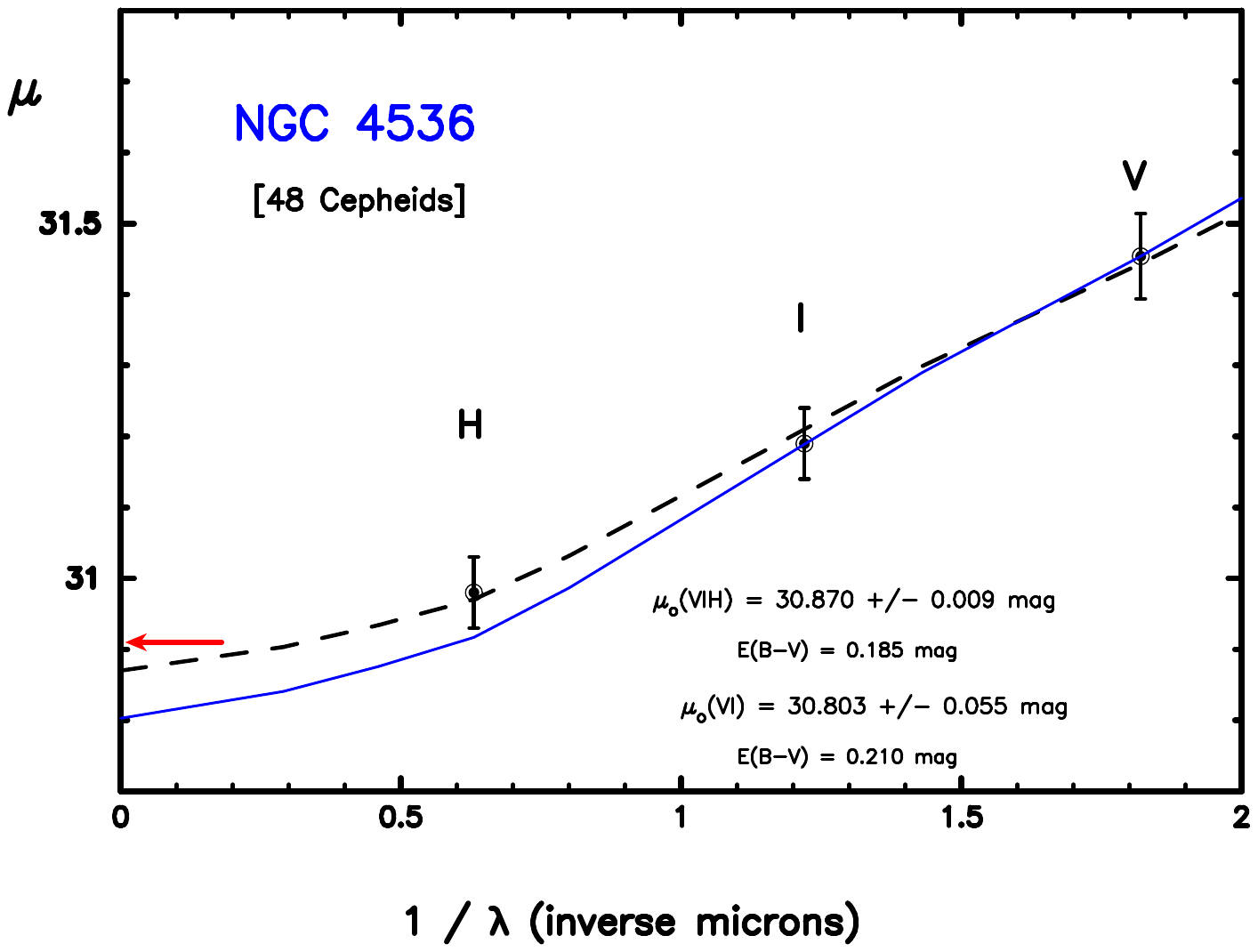}

\includegraphics[width=8.0cm,angle=-0]{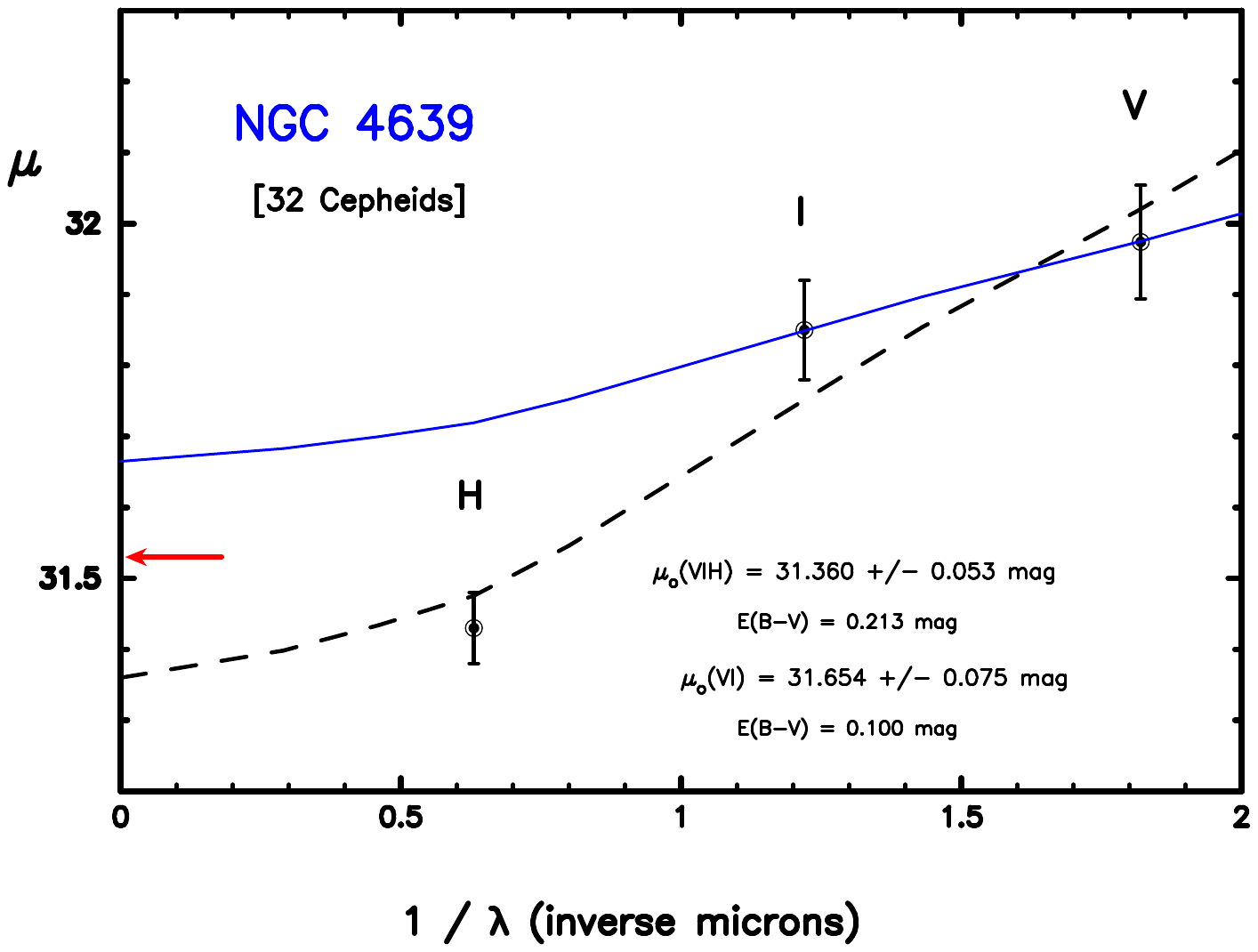}
\includegraphics[width=8.0cm,angle=-0]{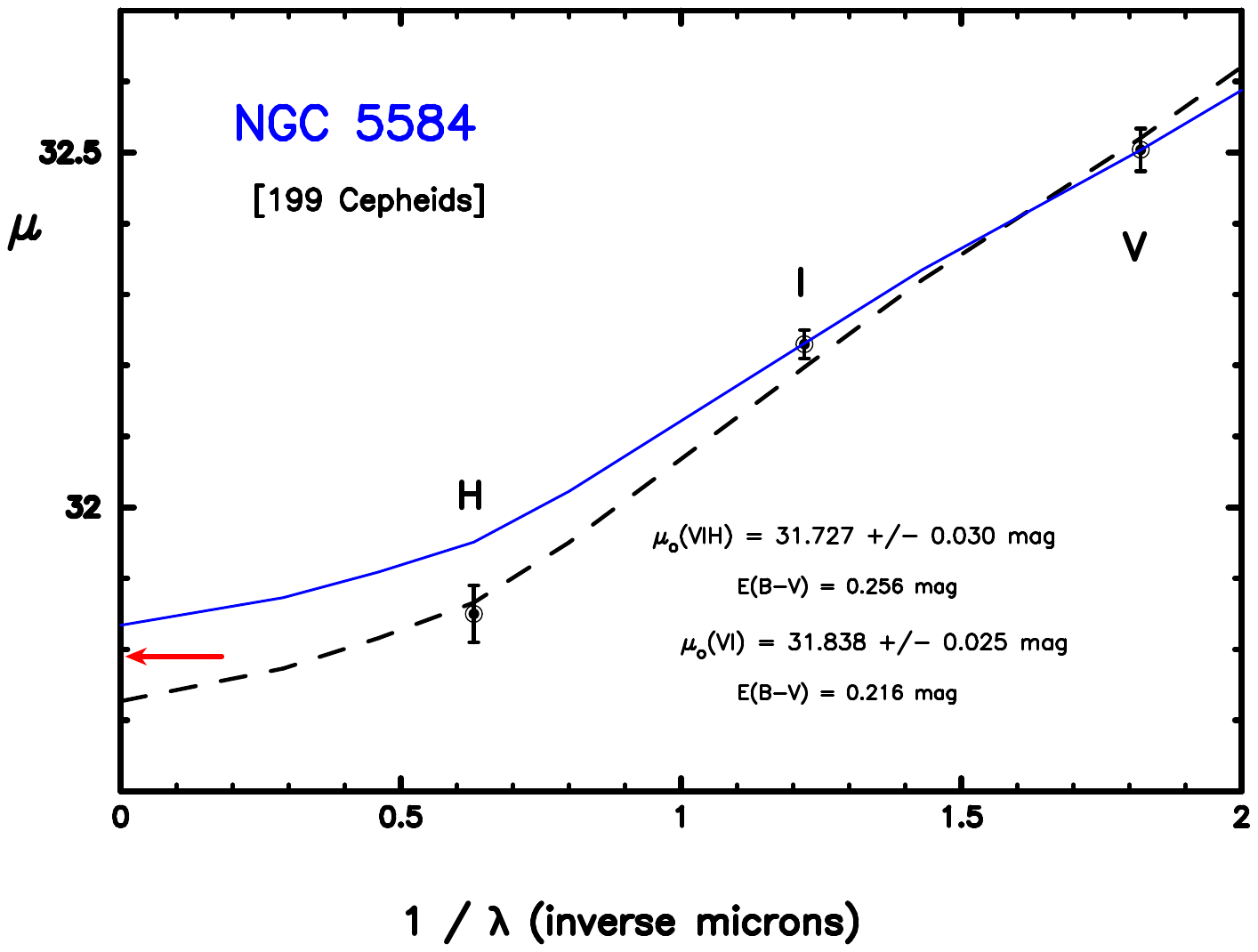}

\includegraphics[width=8.0cm,angle=-0]{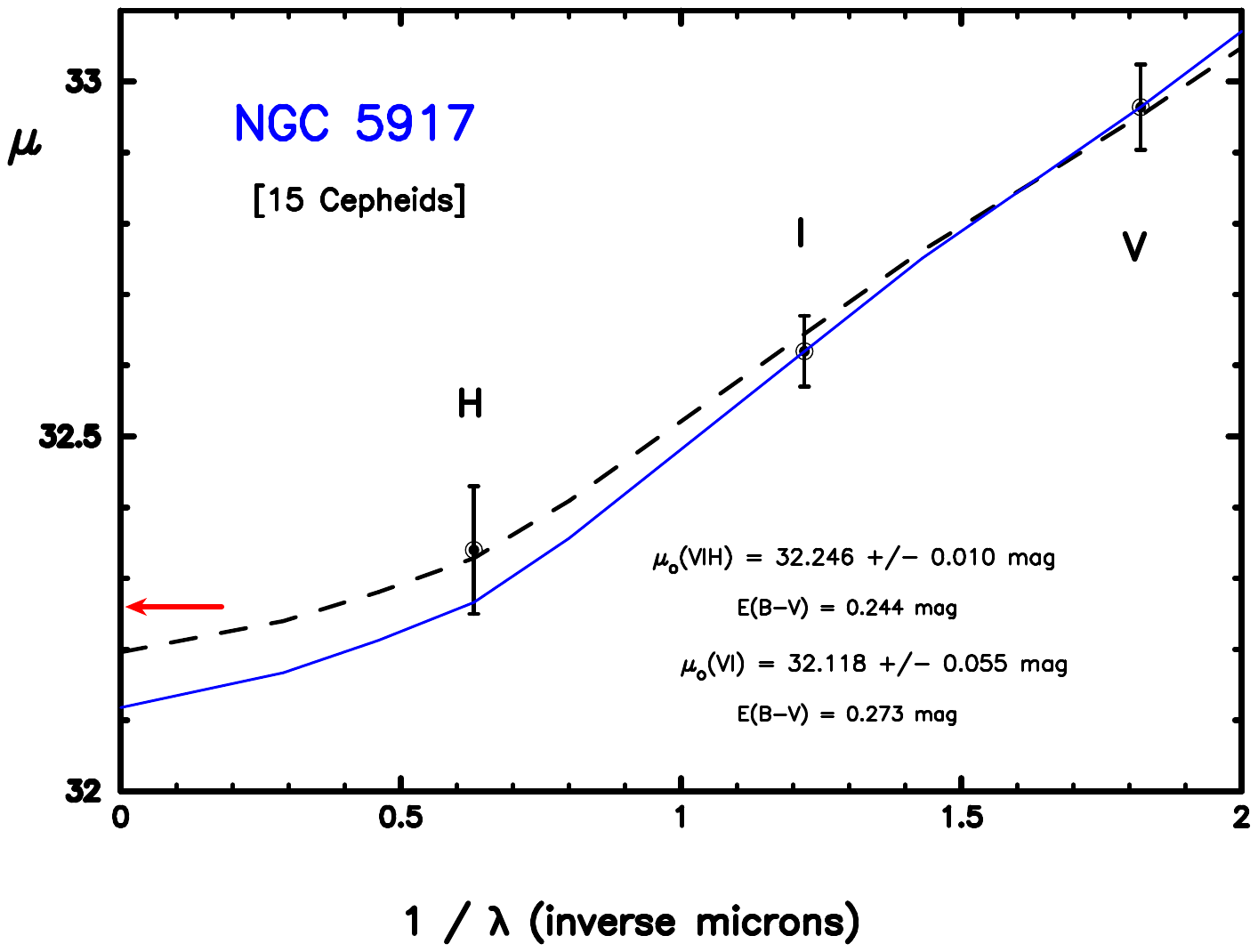}
\includegraphics[width=8.0cm,angle=-0]{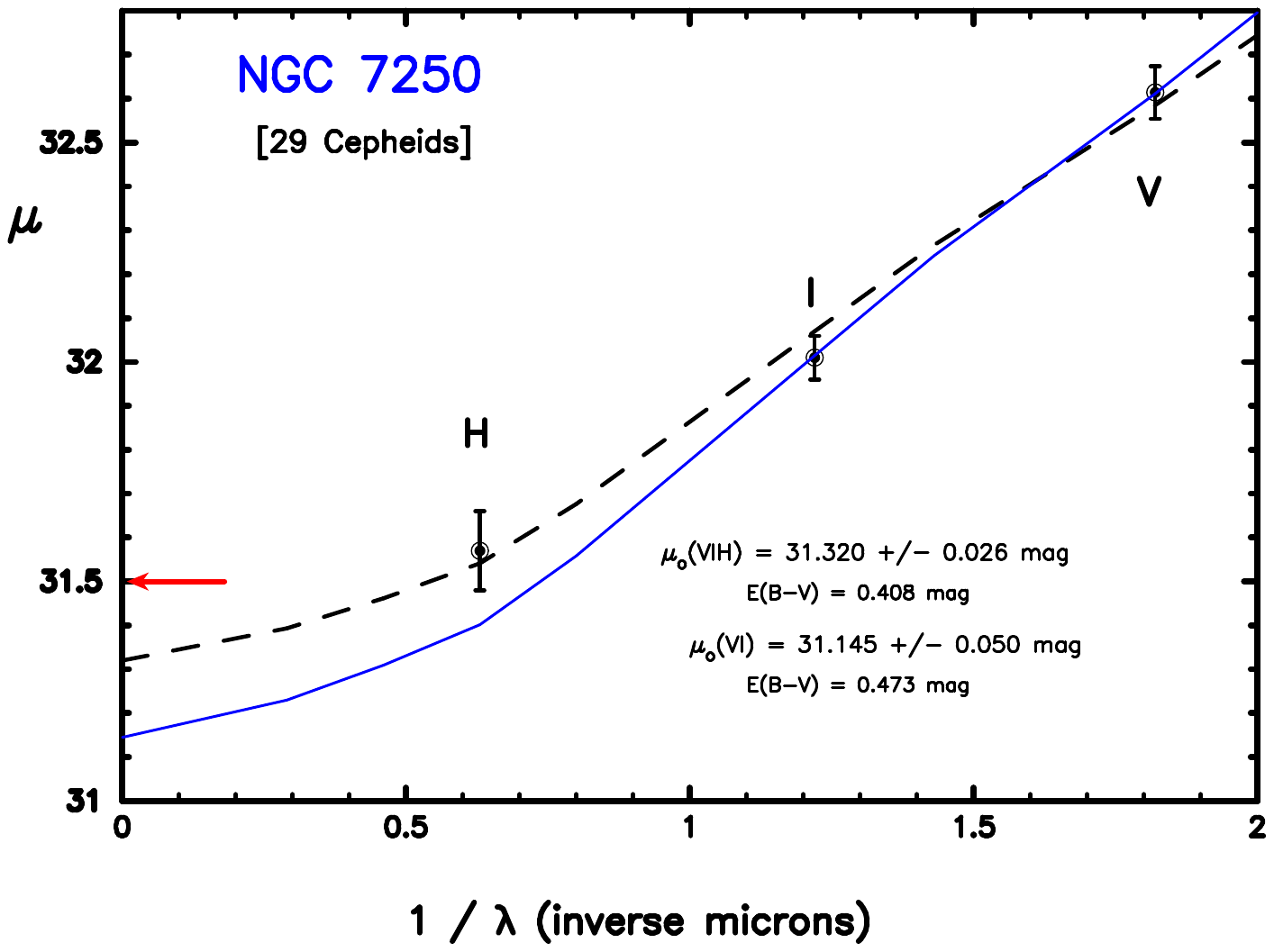}
\caption{\small Same as for Figure 7.}
\end{figure*}

\begin{figure*}
\centering
\includegraphics[width=10.0cm,angle=-0]{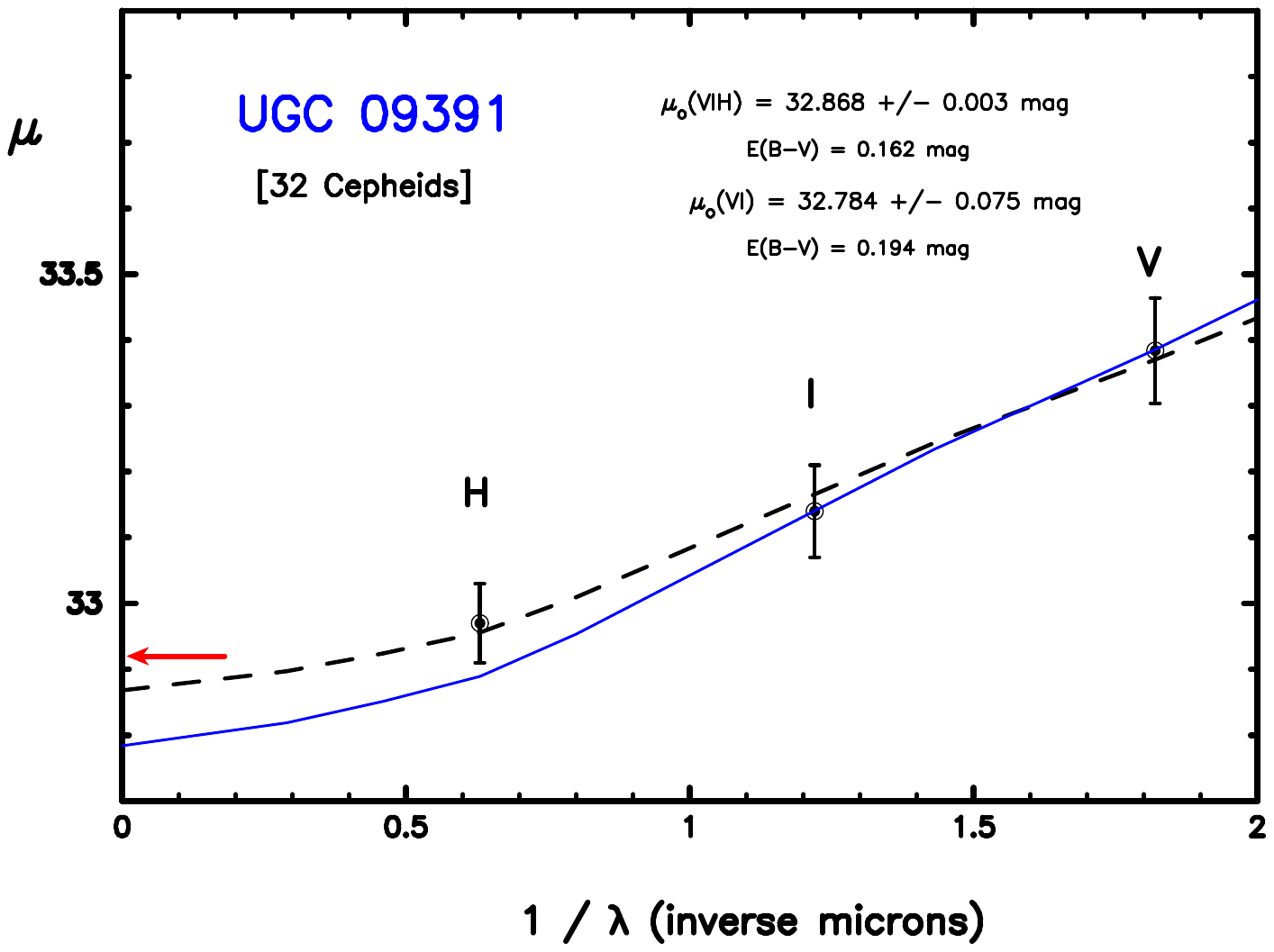}
\caption{\small Same as for Figure 7.}
\end{figure*}

\begin{figure*} 
\centering 
\includegraphics[width=12.0cm, angle=-0]{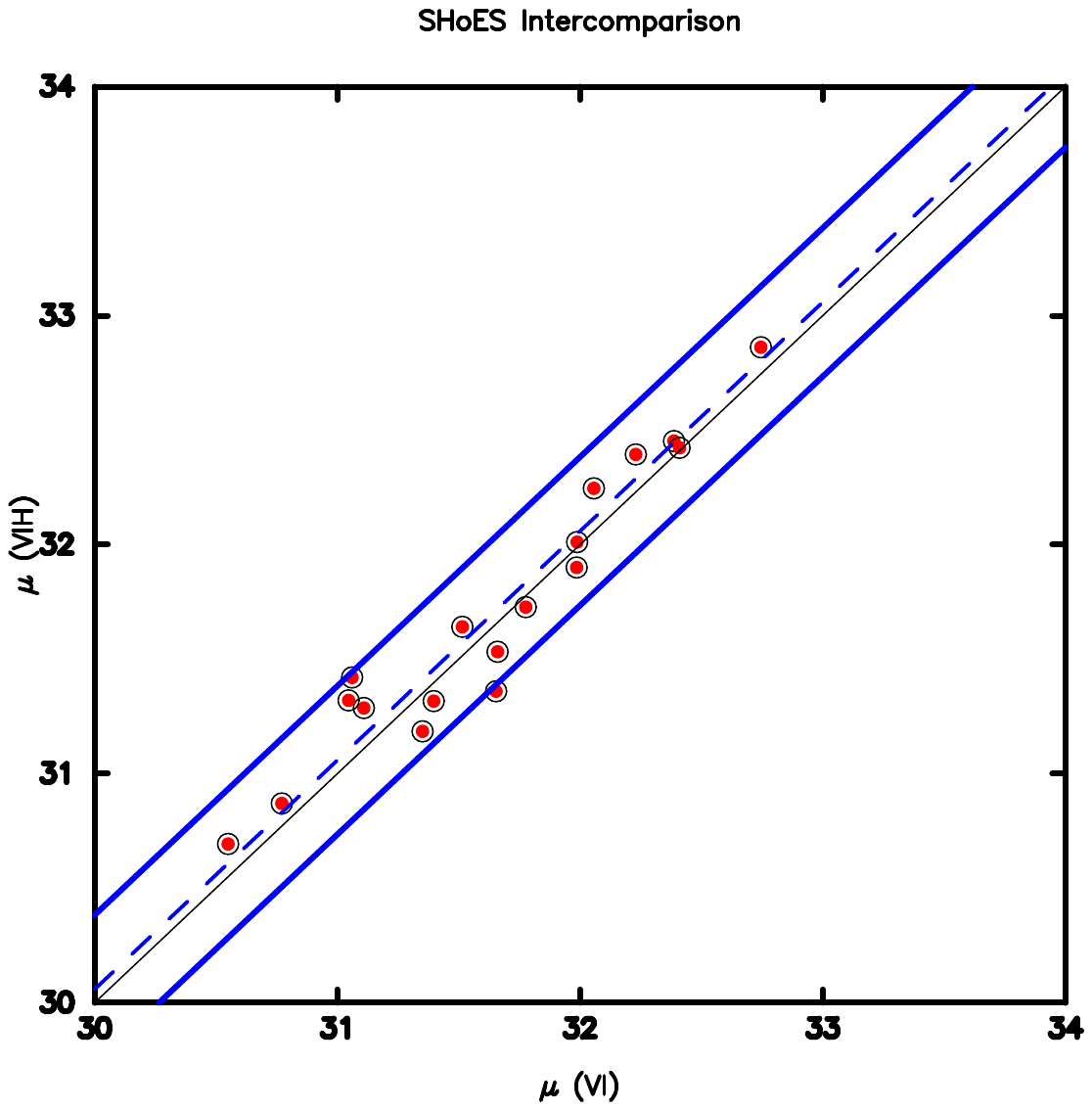} 
\caption{\small Comparison of the W(VIH) and W(VI) distance moduli of SHoES galaxies. 
The W(VIH) moduli are plotted in the vertical direction; the W(VI) moduli are tracked along the horizontal direction. 
The mean offset between the two determinations is shown by the dashed blue line of unit slope. Two-sigma (solid, blue) lines, encompassing the scatter, flank that line and are separated by $\pm 0.32$~mag. 
A thin black line shows the one-to-one correspondence had there been no zero-point offset. }
\end{figure*}

\clearpage
\begin{deluxetable*}{lcccccc}
\tablecaption{Cepheid VIH and VI True Distance Moduli} 
\label{tab:VIH-VI} 
\tablehead{\colhead{Galaxy} & \colhead{$\mu$(VIH)} & \colhead{$\sigma$(mag)} & \colhead{$\mu$(VI)}& \colhead{$\sigma$(mag)} & }
\startdata
NGC 1015 & 32.455  & $\pm$0.017 & 32.499 & $\pm$0.061 &       \\
NGC 1309 & 32.422  & 0.028 & 32.498 & 0.035 &       \\
NGC 1365 & 31.287  & 0.035 & 31.179 & 0.045 &       \\
NGC 1448 & 31.191  & 0.048 & 31.420 & 0.041 &       \\
NGC 2442 & 31.420  & 0.040 & 31.279 & 0.020 &       \\
NGC 3021 & 32.402  & 0.012 & 32.297 & 0.071 &       \\
NGC 3370 & 32.015  & 0.017 & 32.081 & 0.035 &       \\
NGC 3447 & 31.910  & 0.040 & 32.046 & 0.025 &       \\
NGC 3972 & 31.518  & 0.055 & 31.770 & 0.041 &       \\
NGC 3982 & 31.646  & 0.021 & 31.587 & 0.041 &       \\
NGC 4038 & 31.312  & 0.023 & 31.397 & 0.038 &       \\
NGC 4258 & 29.460 & 0.069 & 29.269 & 0.060 &       \\
NGC 4424 & 30.693  & 0.008 & 30.623 & 0.280 &       \\
NGC 4536 & 30.870  & 0.009 & 30.803 & 0.055 &       \\
NGC 4639 & 31.360  & 0.053 & 31.654 & 0.075 &       \\
NGC 5457 = M101 & 28.987  & 0.015 & 28.890 & 0.015 &       \\
NGC 5584 & 31.727  & 0.030 & 31.838 & 0.025 &       \\
NGC 5917 & 32.246  & 0.010 & 32.118 & 0.055 &       \\
NGC 7250 & 31.320  & 0.026 & 31.145 & 0.050 &       \\
UGC 9391 & 32.868  & 0.003 & 32.784 & 0.075 &       \\
\enddata
\end{deluxetable*}


\section{Summary Discussion}
We have revisited the differential (Cepheid comparison with the TRGB) test of the dependence of the Cepheid Leavitt law on metallicity. We find that for an updated sample of 29 nearby galaxies the TRGB and Cepheid distance scales are in good systematic agreement, with a mean offset of --0.034~mag {$\pm$0.007}~mag (error on the mean) at 12 + log(O/H) = 8.50 dex.
We note that this agreement lends added independent support to the value of the TRGB zero point of $M_I = $~--4.05~mag recently derived by Freedman et al. (2019, 2020). The overall scatter in the differences between the paired measurements is $\pm $ 0.099~mag (giving a sigma on the mean of $\pm $0.022~mag.)  
If these errors are equally shared between the two methods, then the errors on the separate Cepheid and TRGB distance moduli are themselves individually good to $\pm$0.070~mag (i.e., $\pm$3.5\% precision in distance). 
However, in Figure 3 we note an abrupt and significant increase in the scatter between the two distance indicators, that being $\pm0.068 $ mag up to a distance modulus of about 30.5~mag (12.5~Mpc) and rising to $\pm0.152$ mag thereafter. This effect was also seen in Figure 5 of Freedman et al. (2019). We do not definitively know the reason for this increased scatter; however, one possibility is crowding and confusion problems in the photometry of the Cepheids in these more distant systems, especially for the high-surface-brightness disk systems (see also the discussion in Wielgorski et al. 2017). 
The mean offset in the zero points of the most nearby distance-selected sample is close to zero, being $\Delta\mu_o$(Cepheid - TRGB) = --0.026 $\pm$~0.015~mag; however, for the more distant sample, there is a larger offset between the two distance scales, amounting to ---0.073 $\pm$~0.057~mag.

Based on a doubling of the sample of galaxies and updated distances used in this study, we find little evidence for any residual correlation of these differential distance moduli with Cepheid-related  metallicities.

\section{Acknowledgements} 
We thank the {\it University of Chicago} and the {\it Observatories of the Carnegie Institution for Science} for their past and on-going support of our long-term research into the calibration and determination of the expansion rate of the Universe. Financial support for this work was provided in part by NASA through grant number HST-GO-13691.003-A from the Space Telescope Science Institute, which is operated by AURA, Inc., under NASA contract NAS 5-26555. 
\normalsize

\section{References}
\par\noindent
Ade, P., Aghanim, N., Arnaud, M.,  et al., 2016, A\&A, 594, 13
\par\noindent
Aghanim et al. 2020 A\&A, 641A, 6P 
\par\noindent
Beaton, R., Seibert, M., Hatt, D., et al.,  2019, \apj, 151, 88
\par\noindent
Bhardwaj, A., Kanbur, S.M., Macri, L.M., et al.,  2016, AJ, 151, 88
\par\noindent
Bono, G. , Caputo, F., Fiorentino, G., et al., 2008, \apj, 684, 102
\par\noindent
Bono, G., Caputo, F., Marconi, M., et al., 2010, \apj, 715, 277
\par\noindent
Bresolin, F., Ryan-Weber, E., Kennicutt, R.C., et al., 2009, \apj, 695, 580
\par\noindent
Breuval, L., Kervella, P., Wielgorski P. et al., 2021, ApJ, 913, 38
\par\noindent
Breuval, L., Riess, A., Kervella, P. et al., 2022, ApJ, 939, 89
\par\noindent
Cerney, W., Freedman, W.L., Madore, B.F., et al. 2020, arXiv: 2012.09701
\par\noindent
Cioni, M.-R., van der Marel, R.P., Loup, C., et al., 2000, A\&A, 359, 601
\par\noindent
da Costa, G.S., \& Armadroff, T.E. 1990, AJ, 100, 162
\par\noindent
Conn, A.R., McMonigal, B., Bate, N.F. 2016, MNRAS, 458, 3282
\par\noindent
Dolphin, A., \& Kennicutt, R.C. 2002, AJ, 123, 207
\par\noindent
Doliphin,A., Saha, A., Skillman, E.D. 2003, AJ, 125, 1261
\par\noindent
Efstathiou, G. 2014, MNRAS, 440, 1138
\par\noindent
Efstathiou, G. 2020, astro-ph 2017.10716
\par\noindent
Freedman, W.L., Madore, B.F., Scowcroft, V. et al., 2012, \apj, 758, 24
\par\noindent
Fernie, D.L. 1969, PASP, 81, 707 
\par\noindent
Freedman, W.L. 1988, ApJ, 326, 691
\par\noindent
Freedman, W.L. 2018, NatAs, 1, 121
\par\noindent
Freedman, W.L., \& Madore, B.F.\ 1990, \apj, 365, 186
\par\noindent
Freedman, W.L. 2021, ApJ, 919, 16
\par\noindent
Freedman, W.L., \& Madore, B.F. 2011, ApJ, 734, 46
\par\noindent
Freedman, W.L., Wilson, C.D., \& Madore, B.F. 1991, ApJ, 372, 455 
\par\noindent
Freedman, W.L., Madore, B.F., Hawley, S.L., et al., 1992, ApJ, 396
\par\noindent
Freedman, W.L., Madore, B.F., Gibson, B., et al., 2001, \apj, 553, 47
\par\noindent
Freedman, W.L., Madore, B.F., Scowcroft, V., et al., 2012, \apj, 758, 24
\par\noindent
Freedman, W.L., Madore, B.F., Hatt, D., et al., 2019, \apj, 882, 34
\par\noindent
Freedman, W.L. \& Madore, B.F. 2023, astro-ph: 2308.02474
\par\noindent
Freedman, W.L., Madore, B.F., Hoyt, T., et al., 2020, \apj, 891, 57 
\par\noindent
Ferraresse, L., Mould, J.R., Kennicutt, R.C., et al., 2000, \apj, 529, 745
\par\noindent
Gibson, B.K, Hughes, S., Stetson, P.B., et al., 1999, \apj, 512, 48
\par\noindent
Gibson, B.K., Stetson, P.B., Freedman, W.L., et al., 2000, \apj, 529, 723
\par\noindent
Gieren, W., Pietrzynski, G., Soszynski, I., et al., 2005, \apj, 628, 695
\par\noindent
Gieren, W., Storm, J., Konorski, P. et al., 2018, A\&A, 620, 99s
\par\noindent
Gorski, M., Pietrzynski, G., \& Gieren, W. et al., 2011, AJ, 141, 194
\par\noindent
Grossi, M., Hwang, N., Corbelli, E. et al. 2011, A\&A, 533, 91
\par\noindent
Hatt, D., Freedman, W.L., Madore, B.F. et al., 2018a, \apj, 861, 104
\par\noindent
Hatt, D., Freedman, W.L., Madore, B.F. et al. 2018b, \apj, 866, 145
\par\noindent
Hoyt, T., Freedman, W.L., Hatt, S., et al., 2019, \apj, 882, 150
\par\noindent
Hoyt, T.J. 2023, NatAs, 7, 590 
\par\noindent
Jacobs, B.A., Rizzi, L., Tully, R.B., et al. 2009, AJ, 138, 332
\par\noindent
Jang, I.-S., \& Lee, M.G. 2017, \apj, 836, 74
\par\noindent
Jang, I.-S., Hatt, D., Beaton, R.L., et al., 2018, \apj, 852, 60
\par\noindent
Johnson, H.L., \& Morgan, W.W. 1953, ApJ, 117, 313
\par\noindent
Kayser, S.E. 1967, Astron. J., 72, 134
\par\noindent
Kanbur, S.M., Ngeow,, C., Nikolaev, N.R., et al., 2003, A\&A, 411, 361
\par\noindent
Kennicutt, R.C., Stetson, P.B., Saha, A., et al., 1998, \apj, 498,181
\par\noindent
Kochanek, C.S. 1997, ApJ, 491,13
\par\noindent
Lee, M-G., Freedman, W.L., \& Madore, B.F., et al., 1993, \apj, 417, 553  
\par\noindent
Madore, B.F., 1982, ApJ, 253, 575
\par\noindent
Madore, B.F. \& Freedman, W.L. 1991, PASP, 103, 933
\par\noindent
Madore, B.F., Freedman, W.L., Owens, K. et al. 2023, AJ (in press) arXiv:230506195M
\par\noindent
McAlary, C.W., Madore, B.F., \& Davis, L.E. 1984, ApJ, 276, 487
\par\noindent
McAlary, C.W., Madore, B.F., McGonegal, R., et al. 1983, ApJ, 273, 539
\par\noindent
McAlary, C.W., \& Madore, B.F. 1984, ApJ, 282, 101
\par\noindent
McGonegal, R., McLaren, R.A., McAlary, C.W. \& Madore, B.F., 1982, ApJL, 257, 33
\par\noindent
Mager, V.A., Madore, B.F., \& Freedman, W.L. 2008, \apj, 689, 721
\par\noindent
Mager, V.A., Madore, B.F., \& Freedman, W.L. 2013, \apj, 777, 79
\par\noindent
Marconi, M., Molinaro, R., Ripepi, V, et al. 2017, MNRAS, 466, 3206
\par\noindent
McQuinn, K.B.W., Boyer, M.L., Mitchell, M.B., et al., 2017, \apj, 834, 78
\par\noindent
Monson, A.J., Freedman, W.L., Madore, B.F. et al., 2012, ApJ, 759, 146 
\par\noindent
Owens, K.A., Freedman, W.L., Madore, B.F., et al. 2022, ApJ, 927, 8
\par\noindent
Pietrzynski, G., Graczyk, D., Gallenne, A., et al., 2006, ApJ, 648, 366
\par\noindent
Pietrzynski, G., Graczyk, D., Gallenne, A., et al., 2019, Nature, 567, 7747 
\par\noindent
Pioto, G., Cappaccioli, M., Pellegrini, C., et al., 1994, A\&A, 287, 371
\par\noindent
Radburn-Smith, D.J., de Jong, R.S., Seth, A.C., et al., 2011, ApJS, 195, 18
\par\noindent
Rich, J.A., Persson, S.E., Freedman, W.L., et al., 2014, \apj, 794, 107
\par\noindent
Riess, A.G., Macri, L.M., Casertano, S., et al., 2011, \apj, 730, 119
\par\noindent
Riess, A.G., Macri, L.M., Hoffmann, S.A., et al., 2016, \apj, 826, 56
\par\noindent
Riess, A.G. Yuan, W., Macri, L.M. et al., 2022, \apj, 934, 7
\par\noindent
Rizzi, L.R. Tully, R.B., Makarov, D., et al., 2007, \apj, 661, 815
\par\noindent
Romaniello, M, Primas, F., Mottini, M., et al., 2006, Mem.S.A.It., 77 172
\par\noindent
Sabbi, E., Calzetti, D., Ubeda, L., et al., 2018, ApJS, 235, 23
\par\noindent
Saha, A.,Thim, F., Tammann, G.A., et al., 2006, ApJS, 165, 185
\par\noindent
Sakai, S., Ferraresse, L., Kennicutt, R.C., et al., 2004, \apj, 608, 42
\par\noindent
Sakai, S., Madore, W.L., \& Freedman, W.L. 1997, \apj, 480, 589 
\par\noindent
Sandage, A.R. 1972, QJRAS, 13, 202
\par\noindent
Scowcroft, V., Freedman, W.l., Madore, B.F., et al. 2013, \apj, 773, 106
\par\noindent
Scowcroft, V. Freedman, W.L., Madore, B.F., et al., 2016, ApJ, 816
\par\noindent
Stetson, P.B., Saha, A., Ferraresse, L., et al., 1998, \apj, 508, 491
\par\noindent
Tikhonov, N.A., Lebedev, V.S., \& Galazutdinova, O.A. et al., 2015, AstL, 41, 239
\par\noindent
Tully, R.B., Courtois, H.M., Dolphin, A., et al., 2013, AJ, 146, 86
\par\noindent
Tully, R.B., Libeskind, N.I., Karachentsev, I.D., et al., 2015, \apj, 802, 25
\par\noindent
van den Bergh, S.,  1975, Stars \& Stellar Systems, Vol.~IX, {\it Galaxies and the Universe}, 
\par ed. A.R. Sandage, M. Sandage \& J. Kristian, (Univ. Chicago Press, Chicago), p. 509
\par\noindent
Udalaski, A., Wyrzykowski, L, Pietryznski, G., et al., 2001, Acta A., 51, 221
\par\noindent
Verde, L., Treu, T., \& Riess, A. 2019, NatAs, 3, 891
\par\noindent
Wielgorski, P., Pietrzynski, G., Gieren, W., et al., 2017, \apj, 842, 116
\par\noindent
Zaritsky, D., Kennicutt, R.C., \& Huchra, J.P. 1994, \apj, 420, 87

\vfill\eject

\appendix
\section{The Wesenheit Function: Definition}

In a few short words, {\it the Wesenheit function is a convenient and mathematically compact means of correcting the apparent magnitude of a star for interstellar extinction.} Its implementation requires a second apparent magnitude at a different wavelength, combined with prior knowledge of the form of the  interstellar extinction curve as a function of wavelength. Figure 12 puts this (implicit) operation into (explicit) graphical form. The equations below describe two views of this same operation, where, in the compact ({\it Wesenheit}) approach, neither the extinction nor the color excess are explicitly solved for; while in the graphical representation, the extinction, in the form of a color excess, is the first quantity to be calculated.

The mathematical expression, in the form of the {\it Wesenheit} function is this: $$W(V,VI) ~~=~~ V - R_{VI} \times (V-I)$$ where $V$ and $I$, in this exemplary case, are apparent magnitudes in the visual and far-red (optical) bands have been chosen, and $R_{VI}$ is the ({\it a priori} known) ratio of total-to-selective absorption, $A_V/E(V-I)$ in those two bands. The latter is empirically defined by curvature in the run of interstellar extinction with inverse wavelength, as shown by the dashed solid line in Figure 12. {\it By construction, the numerical value of $W$ is independent of the amount extinction}, as can be seen in the following lines of unpacking the embedded definitions of various terms:

$$V = V_o + A_V$$
$$I = I_o + A_I$$
$$E(V-I) ~\equiv~ (V-I) - (V-I)_o ~~=~~ (A_V - A_I)$$
$$R_{VI} ~\equiv~ A_V/E(V-I)$$
or equivalently, put in a form that will be used below,
$$A_V - R_{VI} \times E(V-I) ~=~ 0.00~~ [1]$$
Then, since 
$$W ~~\equiv~~ V - R_{VI} \times (V-I)$$
$$W ~~=~~ [V_o + A_V] ~~-~~ [R_{VI} \times ((V-I)_o + E(V-I))]$$  
and then regrouping intrinsic (dereddened) terms on the left and extrinsic (reddening) terms on the right, gives
$$W ~~=~~ [V_o + R_{VI} \times (V-I)_o] ~~-~~ [A_V - R_{VI} \times E(V-I)]$$
where the second term is, by definition, identically zero (as given above in Equation [1]) for all values of the extinction, and so 
$$W ~~=~~ [V + R_{VI} \times (V-I)] ~~=~~ [V_o + R_{VI} \times (V-I)_o] = W_o$$
or 
$$W = W_o$$

\noindent
$W$, formed from apparent magnitudes and apparent colors, is then numerically identical to $W_o$, had it been calculated from intrinsic magnitudes and intrinsic colors. And this occurs without any explicit determination of the extinction or reddening. $W$ {\it is an intrinsic magnitude, by construction.}

\subsection{The Earlier Origins and Conceptual Precursors of W}
\subsubsection{ Johnson \& Morgan (1953): The Reddening-Free Color, $Q$}

It should be noted that $W$, the {\it Wesenheit} function, has a shared lineage with Johnson \& Morgan's (1966) $Q$ function, which is a {\it reddening-independent color}. $Q$ is formed in the following way: using $UBV$ bands for illustration purposes, $(U-B)$ \& $(B-V)$ colors, and $E(B-V) = (B-V) - (B-V)_o$ \& $E(U-B) = (U-B) -(U-B)_o$ color excesses can be combined using $$X_{UBV} \equiv E(U-B)/E(B-V)$$ (which is the ratio of color excesses, standing in for $R_{VI}$ as used above in the formulation of $W$) and then similarly re-grouped as $$E(U-B) - X_{UBV} \times E(B-V) = 0.00 $$ to be used below to cancel their individual reddening effects.

The reddening-free color, $Q$ is defined using apparent (i.e., reddened) colors as 
$$Q ~\equiv~ (U-B) - X_{UBV} \times (B-V))$$
Unpacking the apparent colors by making the color excesses explicit gives

$$Q ~=~ [(U-B)_o + E(U-B)] - X_{UBV} \times [(B-V)_o + E(B-V)]$$

Expanding, and again grouping like terms, gives 
$$Q ~=~ [(U-B)_o - X_{UBV} \times (B-V)_o] ~~-~~ [E(U-B) - X_{UBV} \times E(B-V)] $$

where it is again seen that the right term vanishes (by definition), leaving
$$Q ~=~ (U-B) - X_{UBV} \times (B-V) ~~=~~ (U-B)_o - X_{UBV} \times (B-V)_o ~=~ Q_o $$
or $$Q = Q_o$$

\subsubsection{van den Bergh (1975): The Cepheid PLC (Intrinsic-Color Independent) Version of W}

The {\it Wesenheit} function also has a history of use that is intimately tied to (and in fact inspired by) the Cepheid Period-Luminosity-Color (PLC) relation. 
That insight came from Sandage's influential rewriting of Leavitt's PL relation into its underlying PLC relation bounded by blue and red edges defining the instability strip on a plane in three dimensions 
where, for example 
$$ M_V ~\equiv~ \alpha~logP ~+~ \beta~(B-V)_o + \gamma.$$

\noindent
van den Bergh's (1975) contribution was to cunningly recast {\it this equation} in such a way that is could resemble a Leavitt Law when plotted in two dimensions, i.e., once again as a PL relation. He did this by adopting a value of $\beta$ (as given by Sandage 1959) and transposing the {\it intrinsic color} term $\beta(B-V)_o$ from the right-hand side of the PLC  to the left forming what he chose to call $W$ (unnamed, but certainly neither an acronym or an initialism, but see below) giving $$W_{vdB} ~~=~~ V_o - \beta ~(B-V)_o ~~=~~ \alpha ~log P - ~\gamma $$
This formulation was designed to suppress (intrinsic-color induced) scatter seen in the raw PL relation by adopting the slope $\beta$ for lines of constant period crossing the Cepheid instability strip, as projected into the plane of the color-magnitude diagram. 
van den Bergh's formulation was {\it not explicitly designed or optimized to cancel out reddening or extinction;} however, he did note that his adopted value of $\beta = \Delta V_o / \Delta (B-V)_o$, the slope of lines of constant period as seen in the CMD (again, as published by Sandage 1959) was closely parallel to reddening lines in the same CMD, as characterized by reddening trajectories having a slope of $R_V = A_V/E(B-V)$. 
At that time $\beta$ was thought to have a value of about 2.5 mag/mag, while the ratio of total-to-selective absorption $R_V$ was, and, is still, thought to have a value of about $R_V = $~3.2 mag/mag. 
This was close enough for van den Bergh to correctly point out that in accounting for the intrinsic-color-induced scatter in the PL relations, a large fraction of the physically distinct, line-of-sight extinction/reddening would also be accounted for, thereby further reducing the scatter in the final resulting PL relation.

\subsubsection{Madore (1982): The Reddening-Free Wesenheit Function}

Later, Madore (1982) chose to define a version of W, not in terms of the Cepheid PLC relation, but rather to
optimize it for all types of stars, and have it defined so as to specifically cancel out all of
the reddening and extinction (without explicitly determining either of them) by having $$W_{BFM}
~~\equiv~~ m_{\lambda_1} - R_{\lambda_1,\lambda_2} \times (m_{\lambda_1} - m_{\lambda_2}).$$ In
this formulation $W$ was explicitly given  a name, the {\it Wesenheit} function, which is  German for ``Intrinsic Essence". 
When applied to Cepheids, the {\it Wesenheit} formulation eliminates a systematic error in the distance scale, the total line-of-sight extinction. 
Inverting van den Bergh's logical argument, however, this new definition of W then coincidentally also results in reducing (but not eliminating) the intrinsic-color-induced scatter in the W-PL relation. 
That latter contribution is random in nature if the instability strip is uniformly sampled, and can be averaged over with no residual systematic error. 

It needs to be noted that the {\it Wesenheit} function will be systematically in error if an incorrect value of $R$ is adopted, and also if $R$ is not universal.  
But abandoning universality would lead to systematic errors across {\it all} attempts to correct for reddening, whether they use the {\it Wesenheit} formulation or not. 
Moving to longer wavelengths, as first suggested in McGonegal et al. (1982), alleviates the magnitude of the extinction correction and decreases the relative uncertainty on the ratio of total-to-selective absorption needed to form a {\it Wesenheit} function at those longer wavelengths.

If a single pair of bands can be used to determine a true (extinction/reddening-corrected) distance modulus (as was done in the HST Key Project), the question can then be raised as to what the point would be in obtaining data at any other longer (or even shorter) wavelengths? 
It has long been (not unreasonably) argued that going to longer wavelengths immediately brings one closer to the true modulus, without making any explicit corrections (McGonegal~et al. 1982), given that the impact of extinction naturally falls off with increasing wavelength. 
For instance, if the optical extinction is expected to be small to begin with ($A_V = $ 0.10 mag, for instance) then the extinction in near infrared (in the 2.2$\mu$m $K$ band, say) will be only 0.01~mag; and for many purposes (and at earlier times) that might have been sufficient (e.g., McAlary et al. 1983; McAlary \& Madore 1984; McAlary, Madore \& Davis, 1984). 
However, when attempting to control systematics at the sub-percent level, and when total line-of-sight extinctions are large (or, in the first instance, totally unknown) then one cannot neglect even the scaled-down residual extinction terms at long wavelengths. 

The HST Key Project used a  two-band ($V$ \& $I$) {\it Wesenheit} approach to compactly and efficiently determine true distance moduli using filters and CCD detectors on board HST in space. Just before the launch of HST, the multi-wavelength ($BVRI$) capabilities of CCDs on the ground were  also explored and implemented. 
Fitting a universal interstellar extinction curve to four independently determined $BVRI$ apparent distance moduli was  undertaken by Freedman (1988) for IC~1613, by Freedman, Wilson \& Madore (1991) for M33, and by Freedman et al. (1993) for NGC~300. 
In those papers direct comparisons were made between the {\it Wesenheit} moduli and the multi-wavelength solutions. 

An important conclusion to be drawn from these studies, and from more recent extensions of the extinction-curve fitting technique, including mid-IR, near-IR and optical data (e.g., Monson et al. 2012 for the LMC, Scowcroft et al. 2013 for IC 1613, Rich et al. 2014 for NGC~6822, and Scowcroft et al. 2016 for the SMC), is that the two-band {\it Wesenheit} function provides true moduli that are consistent (to within their mutually quoted uncertainties) with the multi-wavelength solutions. 
It is clear that the limiting factor in any of these applications is the number of Cepheids observed and the precision of the photometry in the individual bands, but not necessarily which band or how many bands were chosen.
Optimization of this process then revolves around telescope aperture, detector wavelength coverage, detector sensitivity and angular resolution/crowding issues. 
With the introduction of new detectors and larger telescopes, in space and on the ground, optimization of the discovery, measurement and dereddening of Cepheids (often at the limits of detection) has become a recurring topic, once again, of renewed interest.



\begin{figure*} 
\centering 
\includegraphics[width=18.0cm, angle=-0]{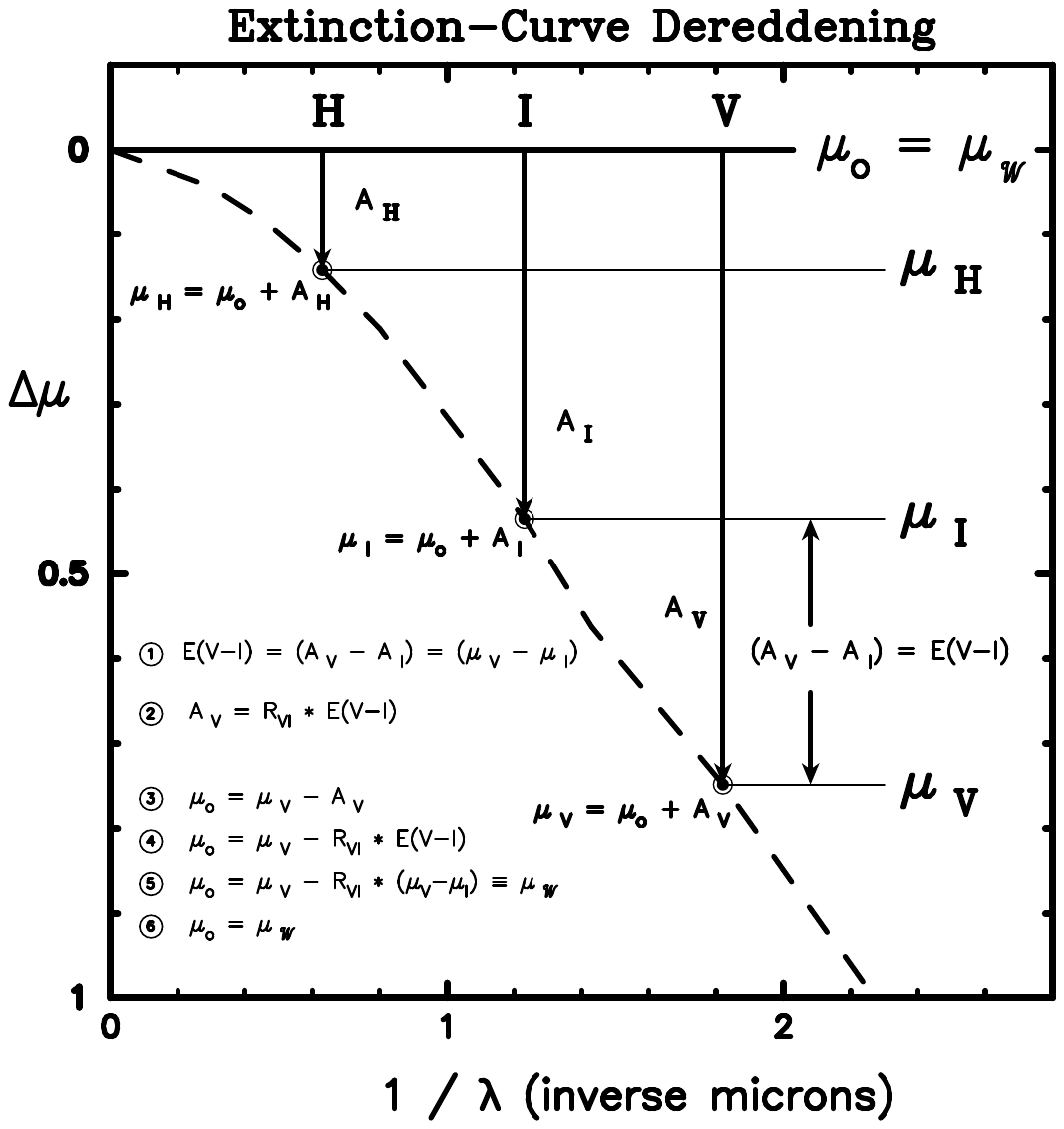} 
\caption{\small A Graphical Representation of the process by which one 
or more apparent distance moduli (circled dots) can be used, in combination to calculate 
a true distance modulus, either by {\it explicitly} calculating the reddening
by differencing apparent distance, turning that difference into an extinction 
and subtracting it from the apparent modulus, or by using the {\it Wesenheit}
function to {\it implicitly} cancel out the reddening and extinction in a simple
one-step formula. The extinction-curve fit is shown by the curved dashed line. The horizontal black line at the top of the plot represents the run of zero extinction at all wavelengths. The difference between that line and the data point is the extinction at that wavelength. The full procedure is described in more detail at the beginning of the Appendix. Furthermore, the sequence of
calculations are numbered and given in the lower left quadrant of this figure, illustrating that $\mu_W$ is equal to $\mu_o$ by definition and construction. }
\end{figure*}
\end{document}